\pdfoutput=1
\documentclass[preprint,12pt]{elsarticle}

\usepackage{xspace,graphicx,mparhack,amsmath}
\usepackage{amssymb,url,underscore,fancyvrb,cancel}
\usepackage{dsfont}
\usepackage{bbm}
\usepackage{picinpar,fancybox}
\usepackage{microtype,relsize}
\usepackage{caption}
\usepackage{comment}
\usepackage{hyperref}
\usepackage{listings}
\usepackage{color}
\usepackage{textcomp}
\definecolor{listinggray}{gray}{0.9}
\definecolor{lbcolor}{rgb}{0.98,0.98,0.98}

\DeclareFontShape{OT1}{cmtt}{bx}{n}{<5><6><7><8><9><10><10.95><12><14.4><17.28><20.74><24.88>cmttb10}{}

\makeatletter
\g@addto@macro\bfseries{\boldmath}
\makeatother

\graphicspath{{./}{figs/}}

\newcommand{\Sherpa} {{\tt SHERPA }\!\!}

\let\oldmarginpar\marginpar
\renewcommand\marginpar[1]{\-\oldmarginpar{\footnotesize \textit{#1}}}

\newcommand{\warnimg}{\includegraphics[height=11mm]{warning}}
\newcommand{\coneimg}{\includegraphics[height=11mm]{cone}}
\newcommand{\bendimg}{\includegraphics[height=11mm]{bend}}
\newcommand{\dblbendimg}{\bendimg\hspace{0.5mm}\bendimg}
\newcommand{\thinkimg}{\includegraphics[height=16mm]{thinker}}

\newcommand{\td}{\ensuremath{\,\textup{d}}}
\newcommand{\Vol}{\ensuremath{\,\textup{Vol}}}

\newcommand{\mccube}{\ensuremath{\mathrm{(MC)}^{3}}}

\begin{document}

\begin{frontmatter}
  
\title{
{\normalsize 
\hspace*{10cm}MCnet-14-10}\\
$\null$ \\
\mccube -- a Multi-Channel Markov Chain Monte Carlo algorithm for phase-space sampling}

  \author[a,b]{Kevin Kr\"oninger}
  \author[a]{Steffen Schumann\corref{author}}
  \author[a]{Benjamin Willenberg}

  \address[a]{II. Physikalisches Institut, Georg-August-Universit\"at G\"ottingen, G\"ottingen, Germany.}
  \address[b]{Lehrstuhl f\"ur Experimentelle Physik IV, Technische Universit\"at Dortmund, Dortmund, Germany.}
  \cortext[author]{Corresponding author.\\\textit{E-mail address:} steffen.schumann@phys.uni-goettingen.de}

  \begin{abstract}
  A new Monte Carlo algorithm for phase-space sampling,
  named \mccube, is presented. It is based on Markov Chain Monte Carlo
  techniques but at the same time incorporates prior knowledge about
  the target distribution in the form of suitable phase-space mappings
  from a corresponding Multi-Channel Importance Sampling Monte
  Carlo. The combined approach inherits the benefits of both
  techniques while typical drawbacks of either solution get ameliorated.
  \end{abstract}


\end{frontmatter}

\noindent

\section{Introduction}
\label{sec:intro}
The stochastic generation of samples from a positive definite
probability density defined over some high-dimensional phase space
poses a challenge in many fields of research. This problem is most
naturally addressed by Monte Carlo methods. In particle physics, Monte
Carlo event generators are used to make theoretical predictions for
the outcome of scattering experiments for example at the Large Hadron
Collider~\cite{Buckley:2011ms}.

Such Monte Carlo generators sample the accessible multi-particle phase
space to generate individual events with a probability density given
by squared transition amplitudes. In turn, fully differential
production rates, corresponding to arbitrary final-state observables,
can be evaluated. Triggered by the experimental needs and theoretical
breakthroughs, amplitude calculations for higher and higher
final-state particle multiplicities became available over the last
years. While tree-level matrix element calculations can be considered
fully automated, see
e.g. Refs.~\cite{Caravaglios:1998yr,Krauss:2001iv,Cafarella:2007pc,
Kilian:2007gr,Gleisberg:2008fv,Alwall:2011uj}, the emerging standard
even for high-multiplicity final states is next-to-leading-order
accuracy in the strong coupling, see
e.g. Refs.~\cite{Berger:2008sj,vanHameren:2009dr,
Hirschi:2011pa,Cullen:2011ac,Cascioli:2011va,Badger:2012pg,Actis:2012qn}. In
practice, the actual calculations are often organized and performed in
event generation frameworks such as {\tt
Helac}~\cite{Bevilacqua:2011xh}, {\tt MadGraph}~\cite{Alwall:2011uj}
or \Sherpa~\cite{Gleisberg:2003xi,Gleisberg:2008ta} that implement the
cross section integration and event generation.

Given the calculational costs for evaluating a complicated $2 \to n$
scattering amplitude, efficient phase-space generation is of utmost
importance. With the high level of optimization that has gone into
amplitude calculations, improvements in phase-space sampling are the
most promising lever arm to further improve existing parton-level
event generators. The existing programs all rely on adaptive Monte
Carlo integration techniques based on {\em Importance Sampling}.  Most
commonly used is the adaptive {\it Multi-Channel Importance Sampling}
(IS) approach
\cite{Papadopoulos:2000tt,Krauss:2001iv,Maltoni:2002qb,vanHameren:2010gg}.
Alternatively, or in combination, methods inspired by
the Vegas algorithm~\cite{Lepage:1977sw} are also
available~\cite{Ohl:1998jn,Jadach:1999sf,Hahn:2004fe,vanHameren:2007pt}.

The construction of suitable phase-space mappings for a Multi-Channel
integrator requires detailed knowledge about the target function to
sample from as it needs to be approximated as good as possible over
the entire phase space. This is typically achieved by analysing the
topology of the contributing scattering amplitudes. There exist known
mappings for essentially all resonance and enhancement structures
occurring in individual topologies. However, due to non-trivial
phase-space restrictions, interference effects or competing
resonances, the multi-channel integrator is never fully efficient. In
particular for processes with many final-state particles these
deficiencies can accumulate and dramatically reduce the overall
sampling efficiency.

An alternative class of algorithms for creating samples according to a
probability density is \emph{Markov Chain Monte Carlo} (MCMC), see
e.g. Ref.~\cite{MCMC} and references therein. These algorithms are
used in a variety of research fields, e.g., astrophysics, biology and
statistical physics. So far, their usage in high-energy particle
physics is rather limited. Examples are the calculation of
cross-sections by integration over multi-dimensional phase
spaces~\cite{Kharraziha:1999iw}, the generation of unweighted events
in next-to-leading order QCD calculations~\cite{Weinzierl:2001ny} or
the mapping of sets of measurements onto high-dimensional parameter
spaces
\cite{Lafaye:2007vs,Lafaye:2009vr}. The most well-known MCMC 
algorithm is the \emph{Metropolis--Hastings
algorithm}~\cite{Metropolis:1953am,Hastings:1970aa}.  A large variety
of advanced algorithms exist, in particular for multimodal problems,
see
e.g. Refs.~\cite{skilling2006,Allanach:2007qj,cappe2008,craiua2009,beaujean2013}. 

In this paper, we present a new sampling algorithm which combines the
prior knowledge of Multi-Channel Importance Sampling with the
flexibility of the Metropolis--Hastings algorithm. We apply the
algorithm to three concrete examples and study the properties of the
samples produced. The first example is an abstract problem and
corresponds to the extreme case where the modelling of the target
function is missing a resonant feature. The second example deals with
the generation of Monte Carlo events for Drell-Yan production serving
as a first illustration of our main application for the new 
algorithm. In our third example we briefly present an implementation
of the algorithm in the \Sherpa event generator and study
its performance in the generation of unweighted events for Drell-Yan
plus multijet production under LHC conditions.  

The paper is structured as follows: we first review the well-known
Multi-Channel Importance Sampling and the Metropolis--Hastings
algorithm in Section~\ref{sec:algorithms} where we also introduce our
new algorithm. Section~\ref{sec:examples} shows three extended examples
for the application of this algorithm, followed by a discussion about
its advantages and disadvantages. The paper concludes in
Section~\ref{sec:conclusions}.

\section{Sampling algorithms}
\label{sec:algorithms}
A standard task in computational physics is the generation of a sample
of random variables distributed according to a normalized target
function $f(x)$. In this section we describe two well-known and
complementary approaches to that problem -- namely, Importance
Sampling Monte Carlo, and in particular Multi-Channel Importance
Sampling, as well as Markov Chain Monte Carlo. We lay out a new method
that attempts to combine the respective advantages of both techniques.

\subsection{Multi-Channel Importance Sampling}

The generation of samples $\{x_{i}\}~(i=1, \dots, N)$ according to a
function $f$ is a byproduct of Monte Carlo integration. Consider the
evaluation of the $d-$dimensional finite integral $\mathcal{I}
= \int_{\Omega_d} f(x) \td x$, with $x \in \mathbbm{R}^d$, over the 
\emph{non-negative} target function $f$ in the integration volume 
$V = \Vol(\Omega_d)$. 
The Monte Carlo estimate for this integral is given through
\begin{equation}
  {\hat{\mathcal{I}}}_N = \frac{V}{N} \sum\limits_{i=1}^N f(x_i) = V \bar{f}\,,
\end{equation}
and an estimate of the corresponding uncertainty is given by
\begin{equation}
  \hat{\sigma}_{\mathcal{I}_N} = V  \frac{\hat{\sigma}_N[f]}{\sqrt{N}}\,,\;\;\text{with}\quad
  \hat{\sigma}_N[f] = \sqrt{\hat{V}[f]} = \sqrt{\frac{1}{N-1} \sum_{i=1}^N(f(x_i)-\bar{f})^2}\, ,
\end{equation}
where $\hat\sigma[f]$ and $\hat{V}[f]$ denote the estimates for the standard 
deviation and variance of $f$, respectively. 

Variance reduction techniques attempt to minimize the uncertainty
estimate by generating a set of points that follow the distribution
$f(x)$ as closely as possible. For a pedagogical introduction see
e.g. Refs.~\cite{James:1980yn,Weinzierl:2000wd}. To illustrate this
technique, we consider a change of the integration variables
\begin{equation}
\mathcal{I} = \int_{\Omega_d} f(x)\td x = \int_{\Omega_d} \frac{f(x)}{g(x)}  g(x) \td x  = \int_{\Omega_d} \frac{f(x)}{g(x)} \td G(x) \, ,
\end{equation}
where $g(x)$ approximates the target function $f(x)$ and is referred
to as \emph{mapping}. We require $g(x)$ to be non-negative,
i.e. $g(x)\geq 0$, and normalizable. We further assume that there
exists an efficient random number generator for samples $\{x_i\}$ according to the
cumulative distribution function $G(x)$, e.g. using the inverse
transformation method \cite{James:1980yn}. It can be shown that for 
suitably chosen $g(x)$ the estimate of the uncertainty of 
$\hat{\mathcal{I}}_N$,
\begin{equation}
  \hat{\sigma}_{\mathcal{I}_N} =
  V\frac{\hat{\sigma}_N\left[f/g\right]}{\sqrt{N}}\, ,
\end{equation}
can be significantly reduced. The resulting set of points $\{x_{i}\}$
is distributed according to the cumulative distribution $G(x)$, and
each point carries a weight $w(x_{i})=f(x_{i})/g(x_{i})$. The weighted
distribution then resembles the function $f$. The events can
be \emph{unweighted} using, e.g., an acceptance-rejection
method~\cite{vonNeumann:1951}.

The outlined method can be generalized to the case where $g(x)$
resembles a sum of $m$ individual mappings, or \emph{channels},
$g_k(x)~(k=1, \dots, m)$, i.e.
\begin{equation}
g(x) = \sum_{k=1}^m \alpha_k g_k(x)\,,\;\;
\text{with}\quad \alpha_k \in [0,1]\;\;\text{and}\;\; \sum_{k=1}^m \alpha_k = 1\,.
\end{equation}
All $g_k(x)$ shall have the same properties as the original function
$g(x)$ discussed above, in particular all cumulants $G_k(x)$ need to
be known and invertible.  The uncertainty estimate for the integral
depends on the choice of the \emph{channel weights} $\alpha_k$.  In
fact, the variance of $f/g$ can be minimized by adapting the channel
weights starting from an initial assignment during a training
integration phase \cite{Kleiss:1994qy}. This decomposition approach is
often referred to as Multi-Channel Importance Sampling.

As typical for an importance sampling technique, it relies on prior
knowledge about the target function $f$ in terms of the approximation
$g$, respectively the individual channels $g_k$. With an optimal
choice for the set of channels the variance of $f/g$ might even vanish
completely; this, however, corresponds to the case that the problem is
solved exactly. In realistic scenarios $f$ can only be approximated
and one is left with a non-vanishing variance. In particular the
unweighting efficiency for generated phase-space points is very
sensitive to the quality of the approximation $g(x)$ to account for
{\em all} pronounced features of the target distribution. Missing a
local structure of $f$ can significantly reduce the global unweighting
efficiency. In practical calculations this can require a huge
proliferation of channels to be considered, rendering their adaptation
and steering computationally challenging. Let us note that, in
contrast to the Markov Chain method discussed next, the phase-space
points $x_i$ generated with Importance Sampling Monte Carlo are
statistically independent, so free of any autocorrelation.

\subsection{Markov Chain Monte Carlo}

An alternative way to generate random numbers according to a target
function $f(x)$ is to use Markov Chain Monte Carlo. This procedure
generates a \emph{Markov Chain}, i.e. a sequence of random values
$X_{k}$ for which the next element, $X_{k+1}$, only depends on the
current state,
\begin{eqnarray}
P(X_{k+1} = x | X_{1}=x_{1}, X_{2}=x_{2}, \dots, X_{k}=x_{k}) = P(X_{k+1} = x | X_{k}=x_{k}) \, .
\end{eqnarray}

A Markov Chain is said to be \emph{time-homogeneous} if $P(X_{k+1} = x
| X_{k}=x_{k})$ is the same for all $k$. A time-homogeneous Markov
Chain can be shown to have a unique stationary distribution
$f(x)$. If, in addition, the Markov Chain is aperiodic, recurrent and
irreducible it is also \emph{ergodic}, i.e. its limiting probability
to reach a certain state does not depend on the initial condition, or
starting value.

The first and probably best known MCMC algorithm is the
Metropolis--Hastings algorithm~\cite{Metropolis:1953am}: for an
arbitrary point $X_{k}=x$, a new point $y$ is generated according to a
proposal function $g(y|x)$. The \emph{acceptance probability} for this
new point is
\begin{eqnarray}
\alpha(y|x)=\min \left( 1, \frac{f(y)g(x|y)}{f(x)g(y|x)} \right) \, .
\end{eqnarray}
Thus, the new point is accepted with a probability of $\alpha$ and so
$X_{k+1}=y$ or, with a probability of $1-\alpha$, the point is
rejected and $X_{k+1}=x$. The conditional probability to move from the
current point $x$ to the point $y$ is \mbox{$p(y|x)=\alpha(y|x)
g(y|x)$} and referred to as \emph{transition kernel},
$\mathcal{K}(y|x)$. For symmetric proposal functions, e.g., a
Gaussian, a flat-top or a Cauchy distribution, the acceptance
probability reduces to $\alpha(y|x) = \min\left(
1, \frac{f(y)}{f(x)}\right)$. Since the acceptance probability
fulfils the requirement of \emph{detailed balance}, i.e.,
\begin{eqnarray}
f(x) p(y|x) = f(x) \mathcal{K}(y|x) = f(y) \mathcal{K}(x|y) = f(y) p(x | y) \, ,
\end{eqnarray}
and the process is ergodic, the resulting sequence is a Markov Chain
with limiting distribution $f(x)$.

Note that the above procedure can introduce an autocorrelation between
points. This can be reduced by introducing a \emph{lag}, i.e. saving
only every $n$th point of the Markov Chain. Also note that reasonable
starting values have to be found for most applications if the number
of samples is small. This can either be done by removing the first few
samples from the Markov Chain (\emph{burn-in phase}) or by running
several chains in parallel and requiring them to mix. Practical
considerations can e.g. be found in Ref.~\cite{Caldwell:2008fw}. An
additional obstacle are multimodal distributions for which the mixing
of several chains and the convergence of a single chain to its
limiting distribution can be poor.

\subsection{Multi-Channel Markov Chain Monte Carlo \mccube}

Importance Sampling and MCMC suffer from complementary
difficulties. In the first case, a bad mapping of the target function
in only one region of phase space can cause a significant drop in
efficiency for the unweighting process. In the latter case, the
samples show an autocorrelation and, in addition, multimodal target
distributions can cause poor mixing or convergence of the Markov
chains.

We propose a combination of both sampling algorithms to
\emph{Multi-Channel Markov Chain Monte Carlo}, or \mccube, which
overcomes the difficulties mentioned above. The algorithm is a variant
of the \emph{path-adaptive Metropolis--Hastings} (PAMH) sampler
proposed in Ref.~\cite{MCMC} in a sense that we use prior analytical
knowledge about the target function in a Metropolis--Hastings
sampler. This information is given by the multi-channel decomposition
used in importance sampling. Technically, \mccube\ mixes two
transition kernels with a common limiting distribution $f(x)$.

The first transition kernel encodes the prior knowledge about the
target function. Let $g_{IS}(x)$ be a probability density which
approximates $f(x)$ corresponding to the multi-channel setup used in
the importance sampling algorithm described earlier. Assume that
efficient random number generators exist for the individual
channels. The proposal function for the first transition kernel,
$g_{IS}(y|x)=g_{IS}(y)$, is generated from $g_{IS}(y)
= \sum_{k=1}^{m} \alpha_k g_k(y)$ and accepted with a probability
\begin{equation}
\alpha_{IS}(y|x) = \min\left( 1,
	\frac{f(y)g_{IS}(x)}{f(x)g_{IS}(y)}  \right) \, .
\end{equation}
The resulting transition kernel is $\mathcal{K}_{IS}(y|x)=\alpha_{IS}
(y|x) g_{IS}( y )$.

The second transition kernel is identical to the one used in the
Metropolis--Hastings algorithm, i.e., $\mathcal{K}_{\mathrm{MH}}(y |
x) = \alpha_{\mathrm{MH}}(y|x) 
g_{\mathrm{MH}}(y|x)$, where
$g_{\mathrm{MH}}(y|x)$ is a symmetric and localized
proposal distribution and the resulting acceptance rate is
\begin{equation}
\alpha_{\mathrm{MH}}(y|x) = \min{\left(1,
	\frac{f(y)}{f(x)}
	\right)} \,.
\end{equation}

In each iteration during the MCMC algorithm, the first (or second)
transition kernel is chosen with a probability of $\beta$ (or
$1-\beta$), with $\beta \in [0,1]$. The combined transition kernel for
the \mccube\ sampler is thus given by
\begin{equation}
\mathcal{K}_{\mccube}(y|x) = 
	\beta  \mathcal{K}_{IS}(y|x) + (1 - \beta) \mathcal{K}_{MH}(y|x).
\end{equation}
The acceptance probability preserves detailed balance since
\begin{eqnarray}
f(x)  \mathcal{K}_{\mccube}(y | x) &= &
	\beta  \min{\left( f(x)  g_{IS}(y|x),
		f(y)  g_{IS}(x|y) \right)} \\ \nonumber
&	&+ (1-\beta)  \min{\left( f(x)  g_{MH}(y|x),
		f(y)  g_{MH}(x|y) \right)} \\ \nonumber
     & = & f(y) \mathcal{K}_{\mccube}(x | y)
\end{eqnarray}
is symmetric in $x$ and $y$. The limiting distribution of the
constructed Markov chain is $f(x)$.

The parameter $\beta$ reflects the confidence of the user in the prior
information to reflect {\em all} relevant features of the target function 
and can be chosen freely. Values close to unity indicate
that the prior knowledge has no uncertainty, i.e. all peak structures
are assumed to be precisely mapped out. In contrast, a value of
$\beta$ equal to zero corresponds to a pure Metropolis--Hastings
algorithm without any prior knowledge.

\subsection{Practical considerations for \mccube}

The \mccube\ algorithm has several free parameters which can be
optimized according to the problem at hand. These are
\begin{itemize}
\item the probability for choosing the IS or MH transition
kernel, $\beta$. This parameter controls to what extend the prior
information is used;
\item the $m$ channel weights, $\alpha_k$, in the
IS transition kernel. These parameters model the decomposition of the
mapping function $g_{\mathrm{IS}}(x)$;
\item the width of the proposal function $g_{\mathrm{MH}}(y|x)$ in 
the MH transition kernel. This parameter has an impact on the sampling
efficiency of the Metropolis--Hastings-part of the algorithm.
\end{itemize}

The optimization procedure followed for the studies presented in this
paper is a sequence of three \emph{pre-runs}: firstly, the channel
weights of the IS transition kernel are optimized by iteratively
updating channel weights. This is accomplished by comparing the target
function with the mapping function using only a small set of samples,
depending on the number of channels employed. Details of this
adaptation strategy can be found in Ref.~\cite{Kleiss:1994qy}. In
consequence, some (very small) channel weights $\alpha_i$ might be
switched off completely, effectively reducing the number of active
channels. Secondly, the width of the proposal function of the MH
transition kernel is adjusted in subsequent sets of samples such that
the efficiency lies within a range of $[0.25, 0.5]$. Thirdly,
the \mccube\ algorithm is run until all chains have converged based on
the $R$-value defined in Ref.~\cite{Gelman:1992zz}. The convergence is
tested on each parameter and the target-function values. The samples
produced during each of the three pre-runs are not saved and thus not
used in the studies. We do not attempt to optimize the parameter
$\beta$ but instead study the properties of the resulting Markov
Chains as a function of $\beta$.

\section{Examples}
\label{sec:examples}
In this section, we present three representative examples for which 
we study the properties and performance of the \mccube\ algorithm. 
The first example simulates the extreme case in which the mapping 
function misses a resonant feature. In contrast, the mapping function 
in the second example approximates the target function rather well 
except for a small region of phase space. In our third example we
present the implementation of the algorithm in the framework of 
the \Sherpa event generator and discuss its application for the 
production of $Z$ plus multijet events under LHC conditions. 

The characteristic measures for assessing the performance of
the \mccube\ algorithm are the \emph{sampling probability}, the number
of calls to the target function, the amount of \emph{autocorrelation}
between the samples and the \emph{convergence} of the Markov Chain to
its limiting distribution. We use several different indicators for
these measures for a predefined number of samples, $N$.

The sampling probability $\eta$ is defined as the number of accepted
points over the number of calls to the target function. In the case of
IS, the sampling probability is equivalent to the unweighting
efficiency. The number of calls to the target function is
$N/\eta$. For the Metropolis--Hastings and \mccube\ algorithms,
the sampling probability is equivalent to the average probability of
changing the state of the Markov Chain in each step. The number of
calls to the target function is $N$.

The autocorrelation factor between subsequent samples $x_{i}$ and $x_{i+1}$
can be calculated for each dimension individually as
\begin{equation}
\rho = \frac{N\sum x_i x_{i+1}-\sum x_i\sum x_{i+1}}
{\sqrt{(N-1)\sum x_i^2-(\sum x_i)^2}~\sqrt{(N-1)\sum x_{i+1}^2-(\sum x_{i+1})^2}} \, .
\label{autocorrelation}
\end{equation}
The autocorrelation is zero if the samples of the Markov Chain are
completely uncorrelated. Values larger than zero are an indication for
sequences of samples with identical states.

The \emph{sequence length} is a measure of the convergence of the
Markov Chain. It is defined as the number of concurrent identical
states in the chain. A poor convergence of the chain can cause a large
amount of autocorrelation and extremely large sequence lengths.

The convergence of the Markov Chain to its limiting distribution can
also be probed by a $\chi^{2}$ discrepancy variable. For a binned
phase space, the number of samples per bin, $n_{i}$, is compared to
the expectation value of the target function in that bin normalized to
the number of samples, $N_{i}$,
\begin{equation}
\label{eqn:chi2}
\chi^2 = \sum_{i} \left(\frac{n_i - N_i}{\sqrt{N_i}}\right)^2 \, ,
\end{equation}
where the sum is over all bins and $\sum_{i}N_{i}
= \sum_{i}n_{i}=N$. In the limit of large numbers, $\chi^{2}$
is distributed according to the well-known
$\chi^{2}$-distribution with a number of degrees-of-freedom equal to
the number of bins. While small $\chi^{2}$ values represent a good
agreement between the sampled distribution and the target function,
large $\chi^{2}$ values indicate a disagreement. It can be
defined over the full phase space or only for a certain fraction.

\subsection{Example one: sampling from a $\Theta$-distribution}

\begin{figure}[!ht]
\centering
\includegraphics[width=0.49\textwidth]{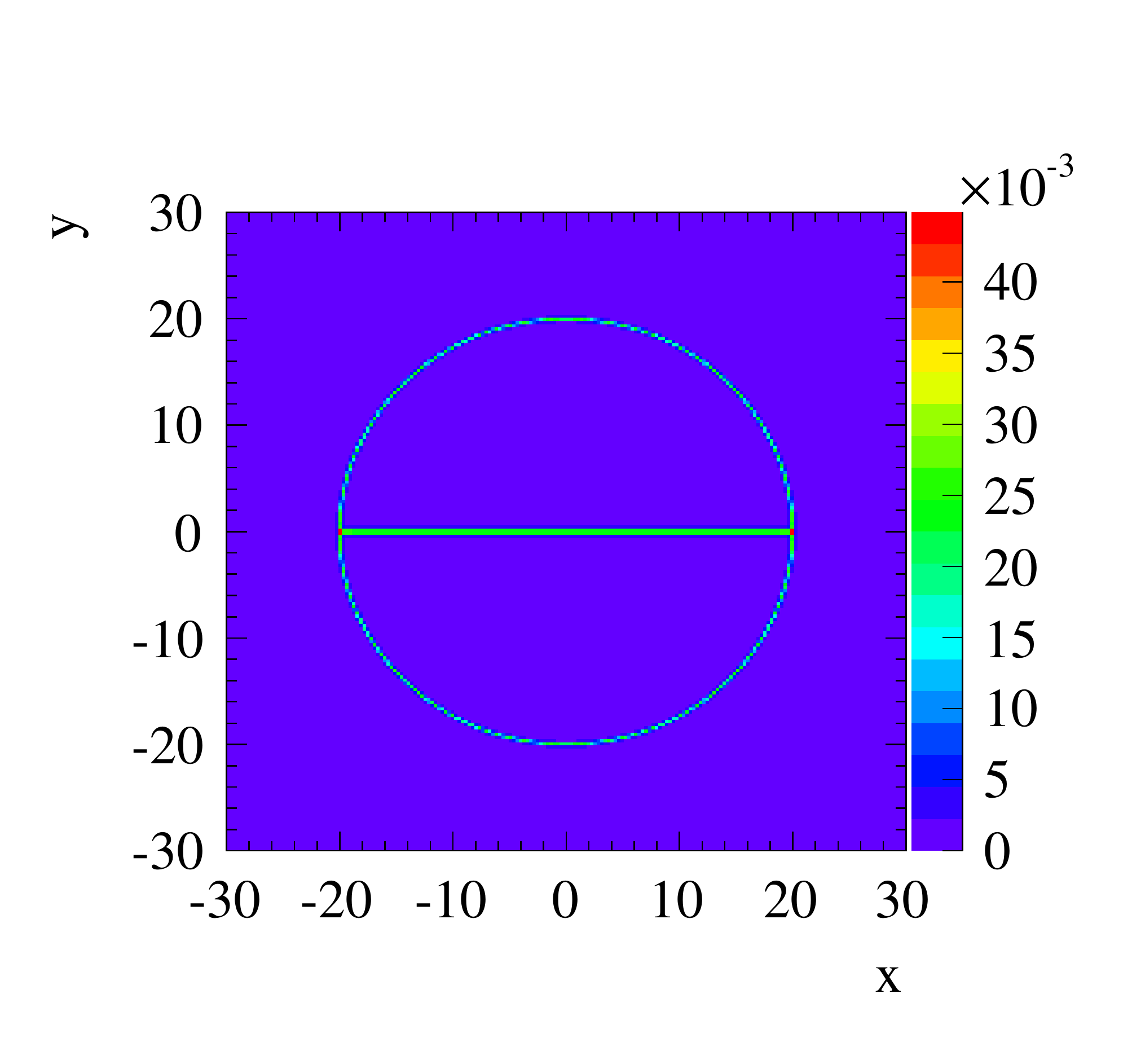}\includegraphics[width=0.49\textwidth]{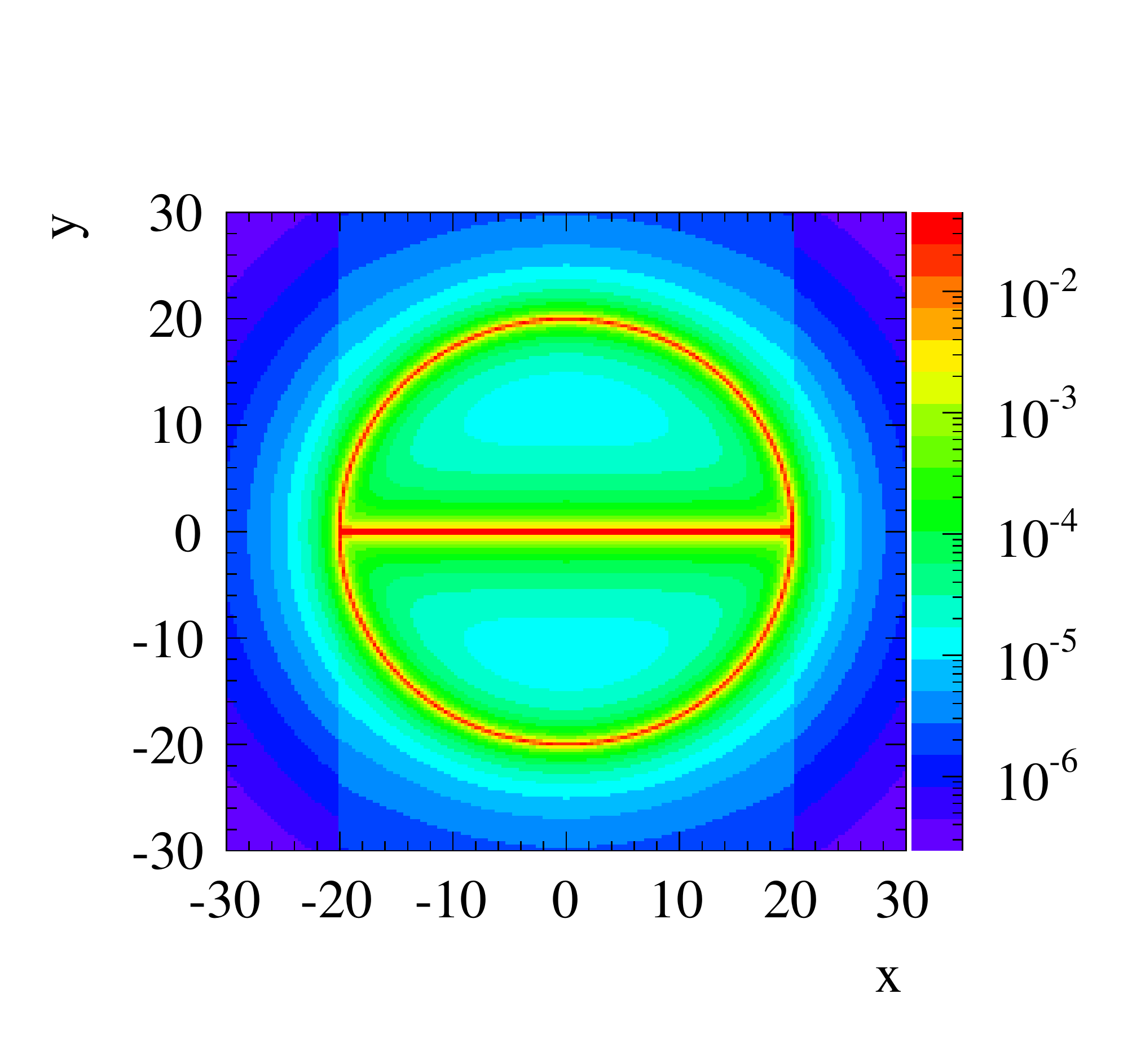}
\caption{The target function $f(x,y)$ with linear (left) and logarithmic (right) $z$-scale.}
\label{fig:thetaDist}
\end{figure}

For this example, we define the target function on
$(x,y) \in \mathbb{R}^2$ as
\begin{eqnarray}
f(x,y) & = & \frac{1}{2 \pi^2} \frac{\Delta r}{(\sqrt{x^2+y^2}-r_0)^2+(\Delta r)^2} \frac{1}{\sqrt{x^2+y^2}} \nonumber \\
& & + \frac{1}{2 \pi  r_{0}} \frac{\Delta r}{(y-y_0)^2+(\Delta r)^2} \theta(r_{0}-|x|) \, ,
\end{eqnarray}
where $\theta(x)$ is the Heavyside function. The shape of the target
function is shown in Figure~\ref{fig:thetaDist}. It resembles the
Greek letter $\Theta$ and it is centred around $x_{0}=y_{0}=0$ with a
radius of the circular part of $r_{0}=20$. The line segment extends
from $x=-r_{0}$ to $x=+r_{0}$ around $y=0$. The circular part and the
line segment have (truncated) Cauchy profiles in $r$ and $y$,
respectively, both with a width parameter of $\Delta r=0.1$. The
resulting distribution is rather narrow.

The mapping function $g(x,y)$ can be split into two channels, a
circular part $g_{1}(x,y)$ and a line segment $g_{2}(x,y)$.

The circular part can best be parameterized in polar coordinates with
radius $r \in \mathbb{R}^+$ and polar angle $\phi \in [-\pi,\pi]$. The
transformation from Cartesian to polar coordinates is denoted $\Phi:
(x,y) \mapsto (r,\phi) = (\sqrt{x^2+y^2},\mathrm{atan}(y/x))$, and the
modulus of the corresponding Jacobi determinant is $|J_{\Phi}| =
r$. Random numbers are generated according to
\begin{eqnarray}
\tilde{g}(r,\phi) = \frac{1}{2\pi^2} \frac{\Delta r}{( r - r_{0})^{2} + (\Delta r)^{2}} \, ,
\end{eqnarray}
i.e. a truncated Cauchy distribution in $r$ in the interval
$[0, \infty]$ centred around $r_{0}$ with a width parameter of
$\Delta r$, and a uniform distribution in $\phi$ in the interval
$[-\pi, \pi]$. The transformation into Cartesian coordinates is
obtained via
\begin{eqnarray}
g_{1}(x,y) = \tilde{g}_{1}(\Phi(x, y)) \frac{1}{|J_{\Phi}|} \, .
\end{eqnarray}

The line segment is parameterized in Cartesian coordinates. Random
numbers are generated according to a uniform distribution in $x$ in
the interval $[-r_{0}, r_{0}]$, and a Cauchy distribution in $y$
centred around $y_{0}=0$ with a width parameter of $\Delta r$. The
parametrization of the second channel is thus
\begin{eqnarray}
g_{2}(x,y) = \frac{1}{2 \pi  r_{0}} \frac{\Delta r}{(y-y_0)^2+(\Delta r)^2} \theta(r_{0}-|x|) \, .
\end{eqnarray}

We study two cases in the following. In the first one, the mapping
function consists of the two contributions defined above,
i.e. \linebreak \mbox{$g(x,y)=\frac{1}{2} g_{1}(x,y)+\frac{1}{2}
g_{2}(x,y)$}. This overall mapping function is identical to the target
function, i.e. $g(x,y) \propto f(x,y)$, up to a global scaling factor,
thus the prior knowledge of the target function is complete. In the
second case, only the circular part is considered in the mapping
function, i.e. $g(x,y)=g_{1}(x,y)$. The model misses a resonant
feature and the prior knowledge of the target function is thus
incomplete. We compare the measures defined for both cases
using IS and the proposed \mccube\ algorithm based on 20 runs of five
chains with 5M samples each. This is done for different values of
$\beta$, in particular $[0,0.01,0.1,0.25,0.5,0.75,0.9,0.99,1.0]$, and
the lag, $[1,2,5,10,20]$. The sampling efficiency for a pure MH
transition kernel, i.e. $\beta=0$, is adjusted to 30\% during the
pre-run.

\subsubsection{Sampling efficiency}

\begin{figure}[!ht]
\centering
\includegraphics[width=0.49\textwidth]{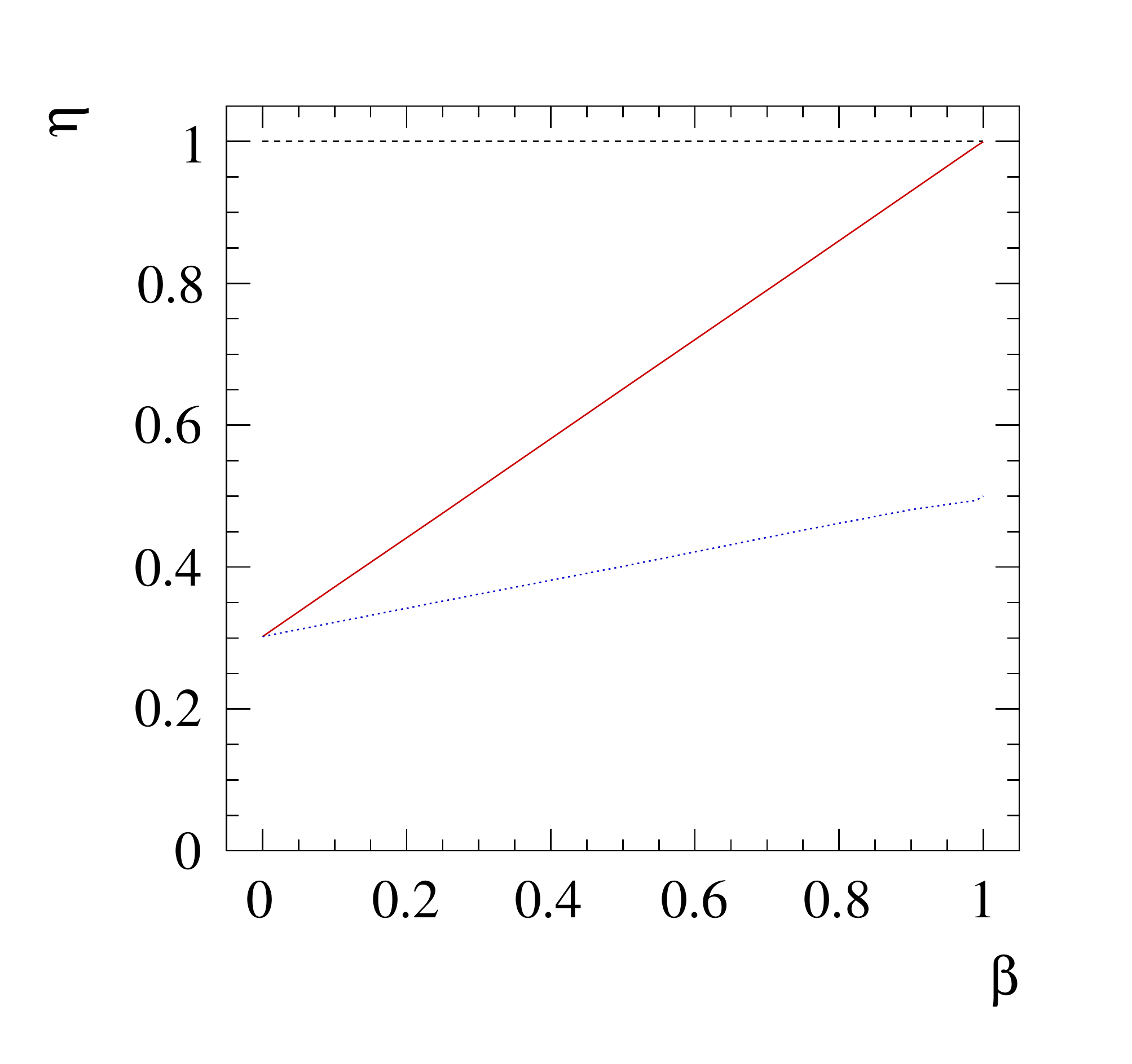}
\caption{The sampling efficiency as a function of $\beta$ for the \mccube\ algorithm for complete (solid red) and incomplete prior knowledge (dotted blue). The 100\% sampling efficiency for the IS algorithm is indicated as a dashed line.}
\label{fig:effEx1}
\end{figure}

Figure \ref{fig:effEx1} shows the sampling efficiency for the \mccube\
algorithm as a function of $\beta$. Since the transition kernel is a
linear combination of two kernels, the efficiency increases linearly
from 30\% at $\beta=0$ to $100$\% ($50$\%) at $\beta=1$ for the case
of complete (incomplete) prior knowledge. In comparison, the sampling
efficiencies for pure IS are $100$\% and $0.01$\%,
respectively. Consequently, a pure IS with incomplete knowledge leads
to factors of up to $5,000$ more calls to the target function compared
to the \mccube\ algorithm.

\subsubsection{Autocorrelation}

\begin{figure}[!ht]
  \centering
  \includegraphics[width=0.49\textwidth]{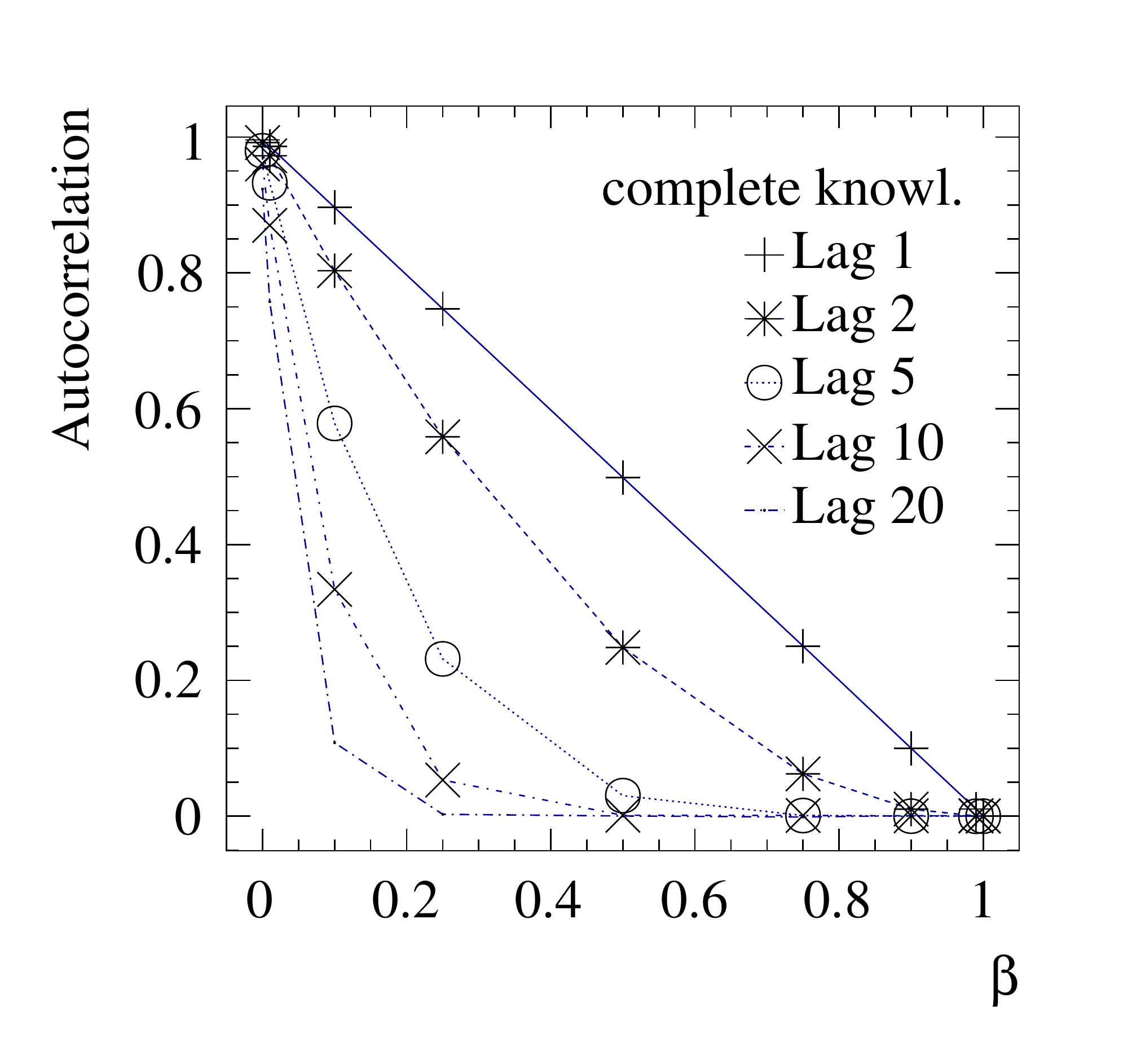}
  \includegraphics[width=0.49\textwidth]{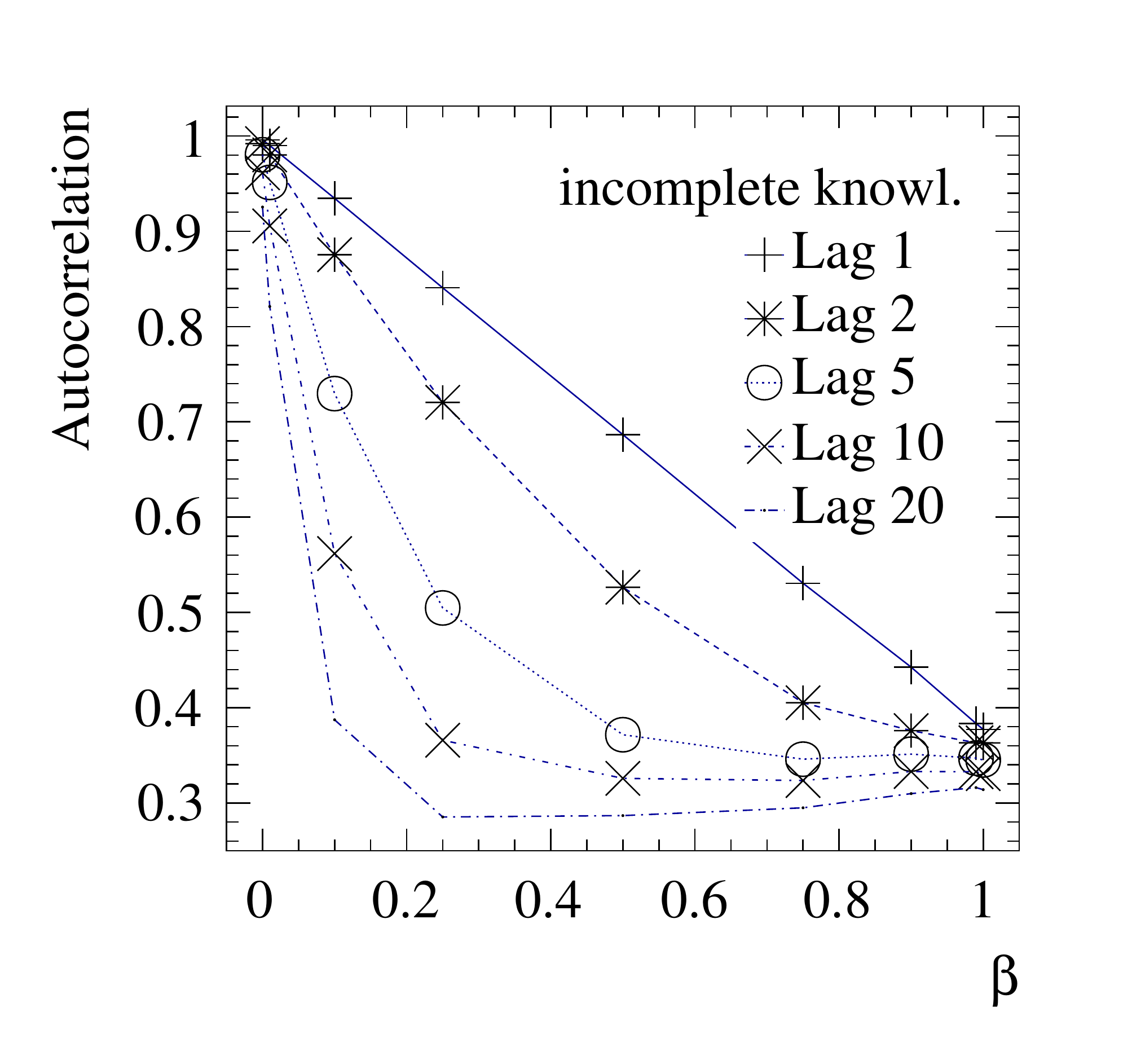} 
  \caption{Autocorrelation for $x$ as a function of $\beta$ for the
  case of complete (left) and incomplete knowledge (right) and for
  different lags.}
\label{fig:autocorrEx1}
\end{figure}

As an example, the autocorrelation for $x$ is shown in
Figure~\ref{fig:autocorrEx1} for the case of complete (left) and
incomplete knowledge (right). For the former case, the autocorrelation
for $\beta=1$ is zero for all lags. This is expected since the
function $f/g$ sampled from is uniform. For admixtures of the MH
transition kernel, the autocorrelation increases to values of 95\% or
larger for $\beta=0$ and lags between 1 and 20. This large
autocorrelation is owed to the small width of the proposal function
used in the MH transition kernel in comparison with the size of
$r_{0}$. As expected, the autocorrelation decreases with an increasing
lag. A similar behaviour is observed for the case of incomplete
knowledge with the exception that the autocorrelation does not reach
zero but a plateau of around 35\%.

\subsubsection{Sequence length}

\begin{figure}[!ht]
	\centering
        \includegraphics[width=0.49\textwidth]{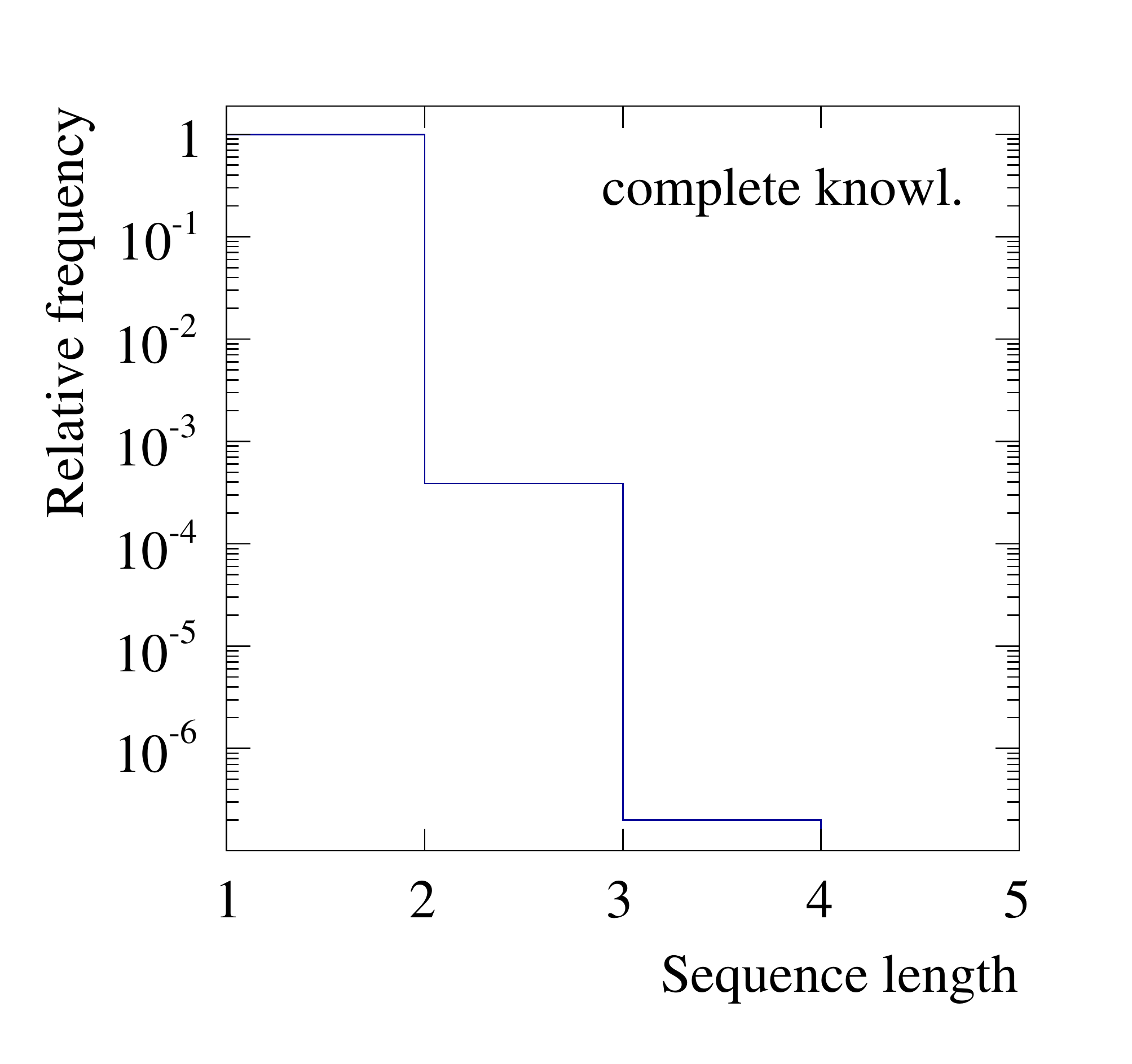}
        \includegraphics[width=0.49\textwidth]{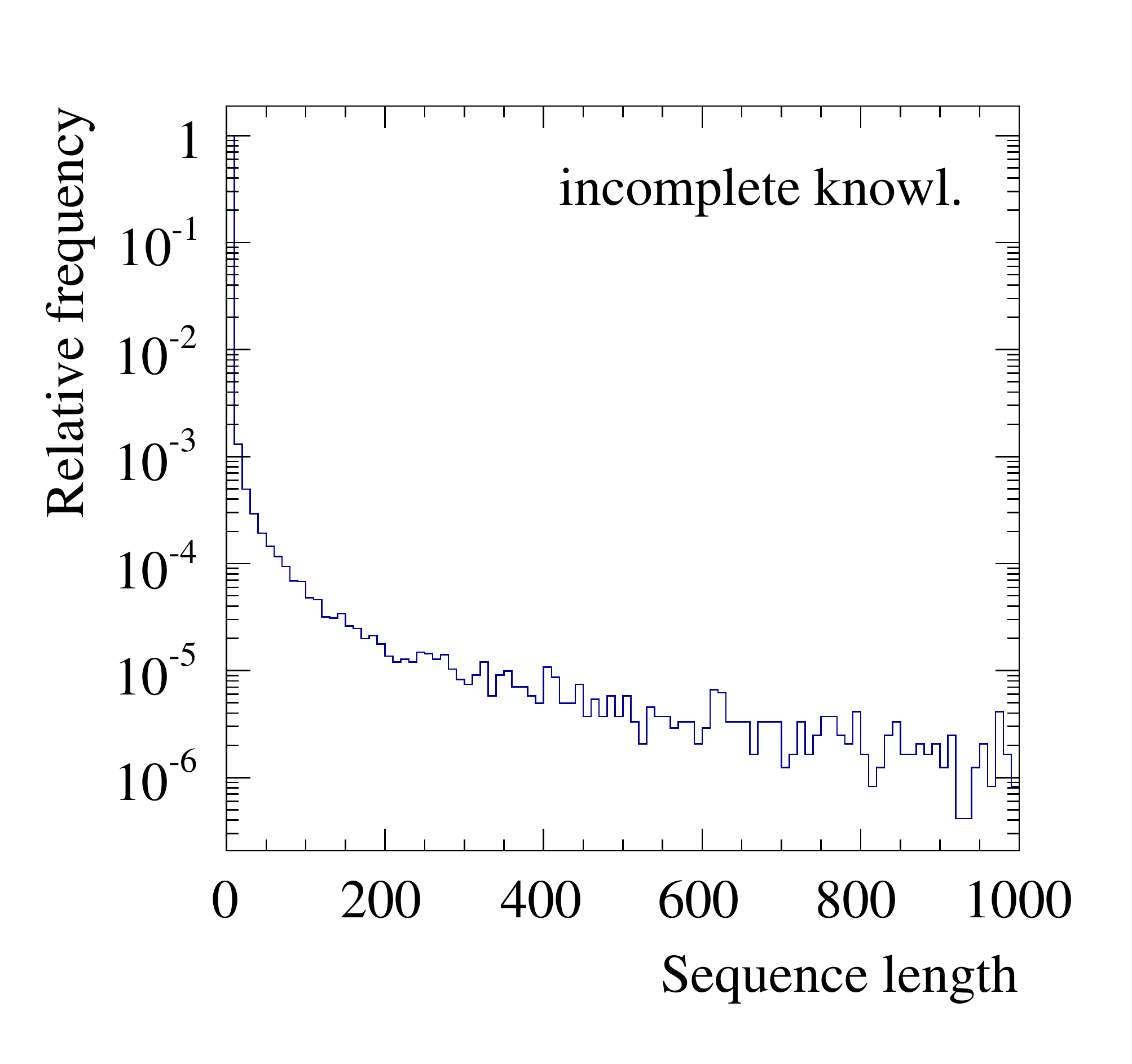}
	\caption{Sequence lengths for the cases of complete (left) and
	incomplete prior knowledge (right) for a lag of one and
	$\beta=1$.}
	\label{fig:ex1SequenceLengthEx}
\end{figure}

The sequence lengths for the cases of complete and incomplete prior
knowledge and $\beta=1$ are shown in
Figure~\ref{fig:ex1SequenceLengthEx} for a lag of one and
$\beta=1$. As expected from the low autocorrelation, the typical 
sequence length in the case of complete prior knowledge is greater than one in
less than a per mil of all cases. In contrast, sequences in the case
of incomplete prior knowledge can reach lengths of greater than~1,000,
although the most likely length is also one.

\begin{figure}[!ht]
	\centering
        \includegraphics[width=0.49\textwidth]{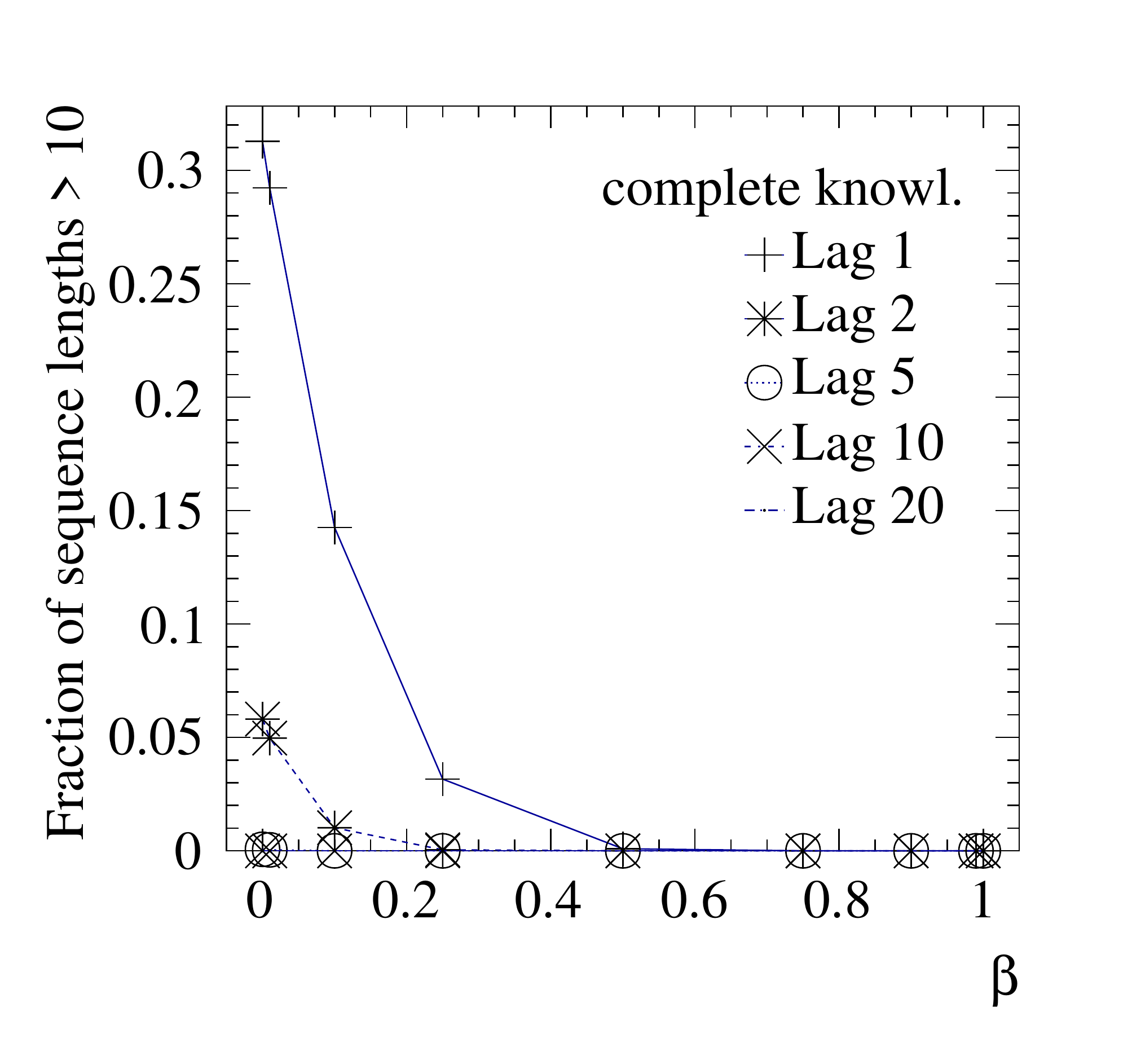}
        \includegraphics[width=0.49\textwidth]{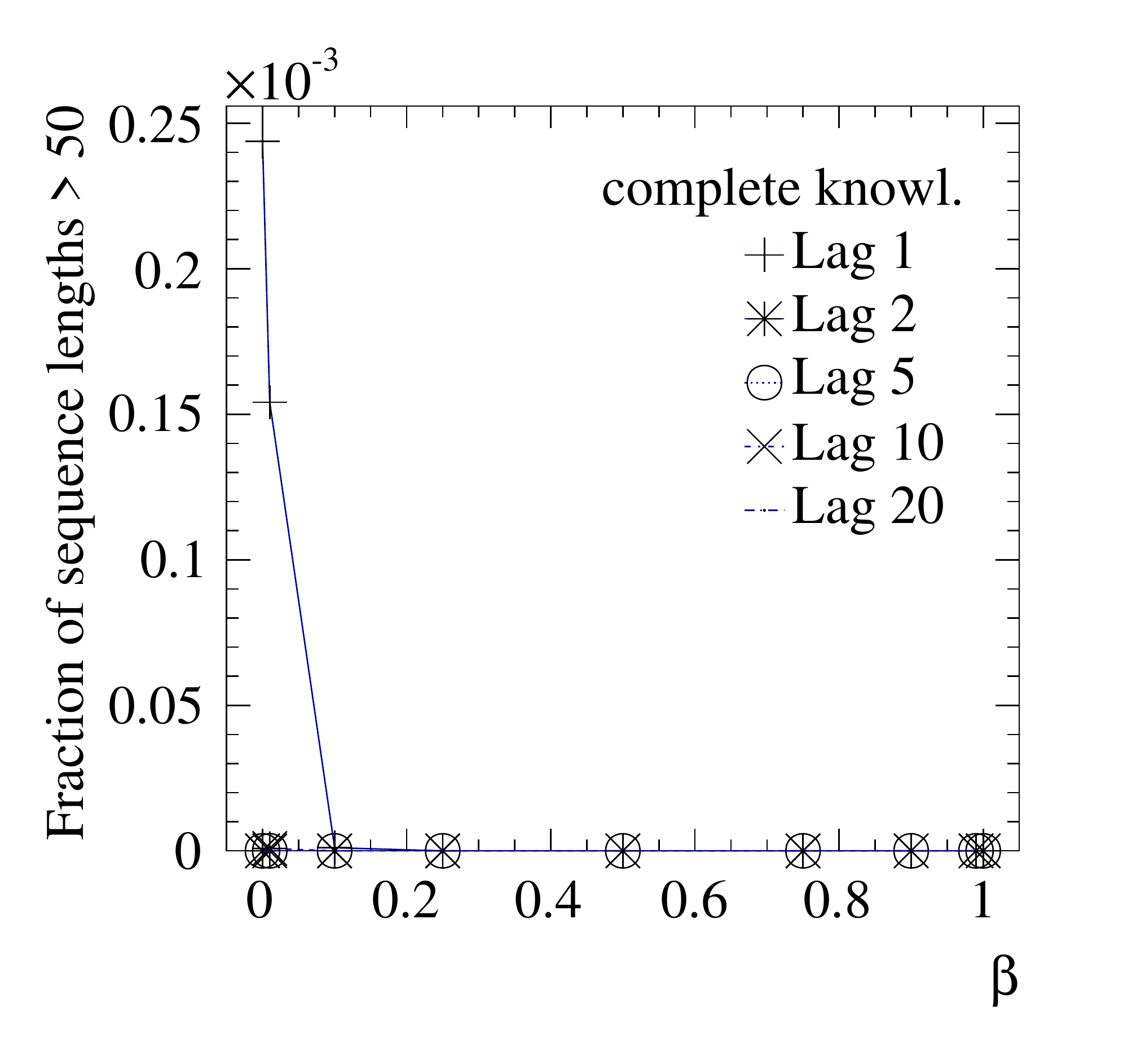} \\
        \includegraphics[width=0.49\textwidth]{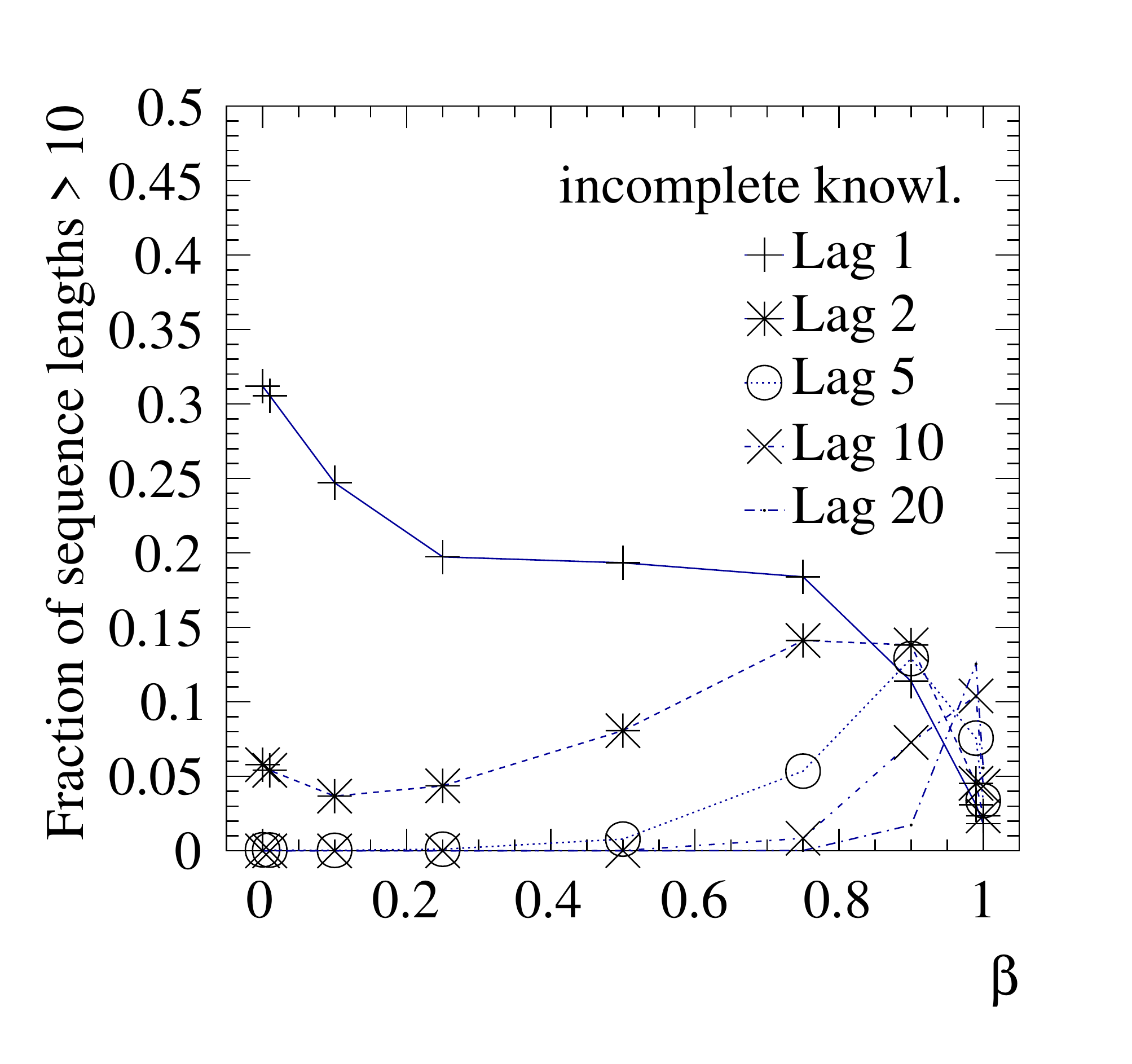}
        \includegraphics[width=0.49\textwidth]{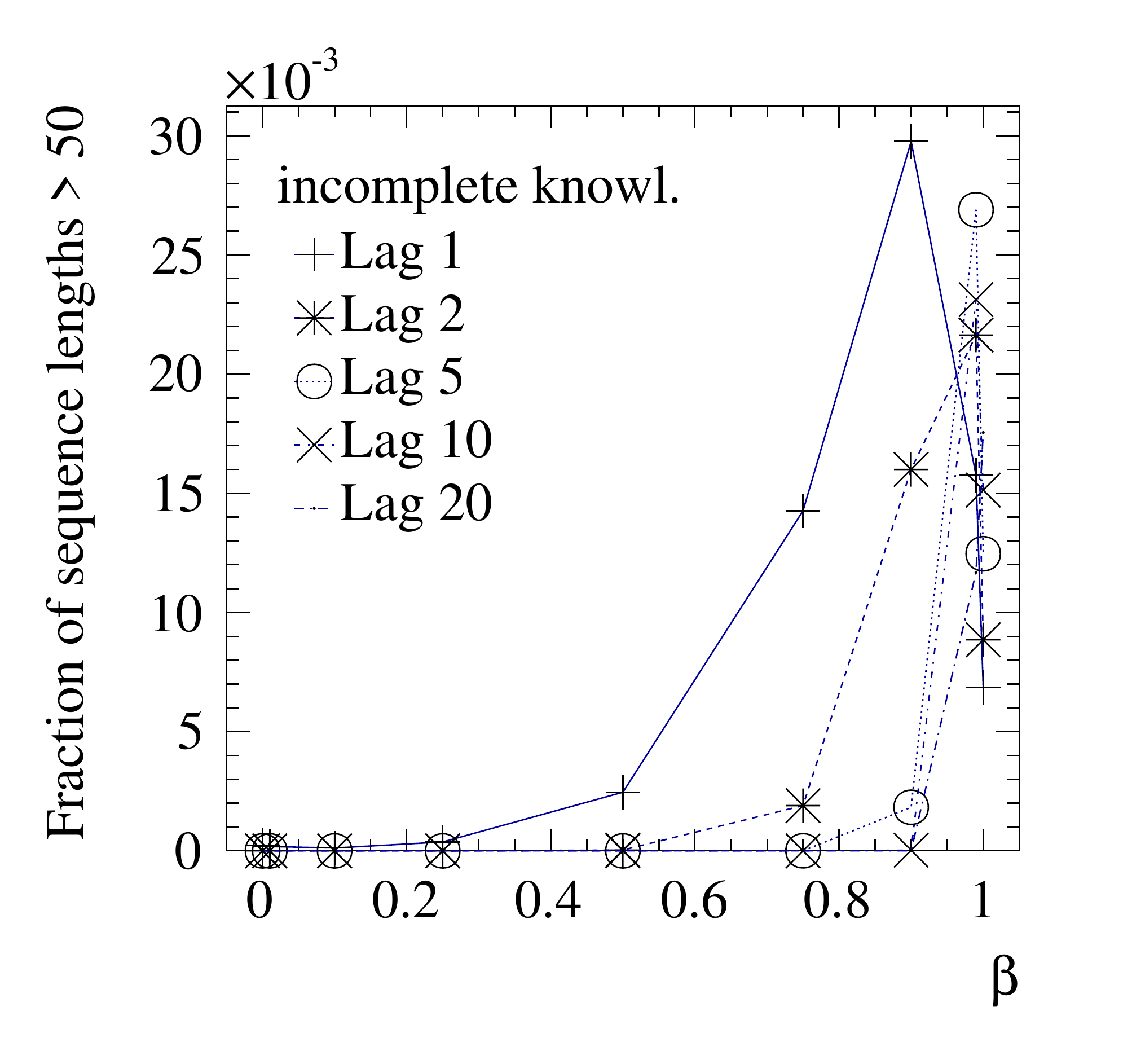}
	\caption{The fraction of sequence lengths greater than 10
	(left) and 50 (right), for the case of complete knowledge
	(top) and incomplete knowledge (bottom), shown for lags
	between 1 and 20.}
	\label{fig:ex1SequenceLength}
\end{figure}

Figure~\ref{fig:ex1SequenceLength} shows the fraction of sequence
lengths above 10 and 50 for the cases of complete and incomplete prior
knowledge. As expected, these fractions drop in the former case with
increasing beta and increasing lag. Both trends can be explained by
the autocorrelation of the MH transition kernel. The fractions are
typically smaller than in the case of incomplete knowledge for which
the fractions increase with increasing beta and decrease with
increasing lag. The former trend is due to the fact that the Markov
Chains get stuck more often if the admixture of incomplete prior knowledge
increases.

\subsubsection{Convergence}

\begin{figure}[!ht]
	\centering
	\includegraphics[width=0.49\textwidth]{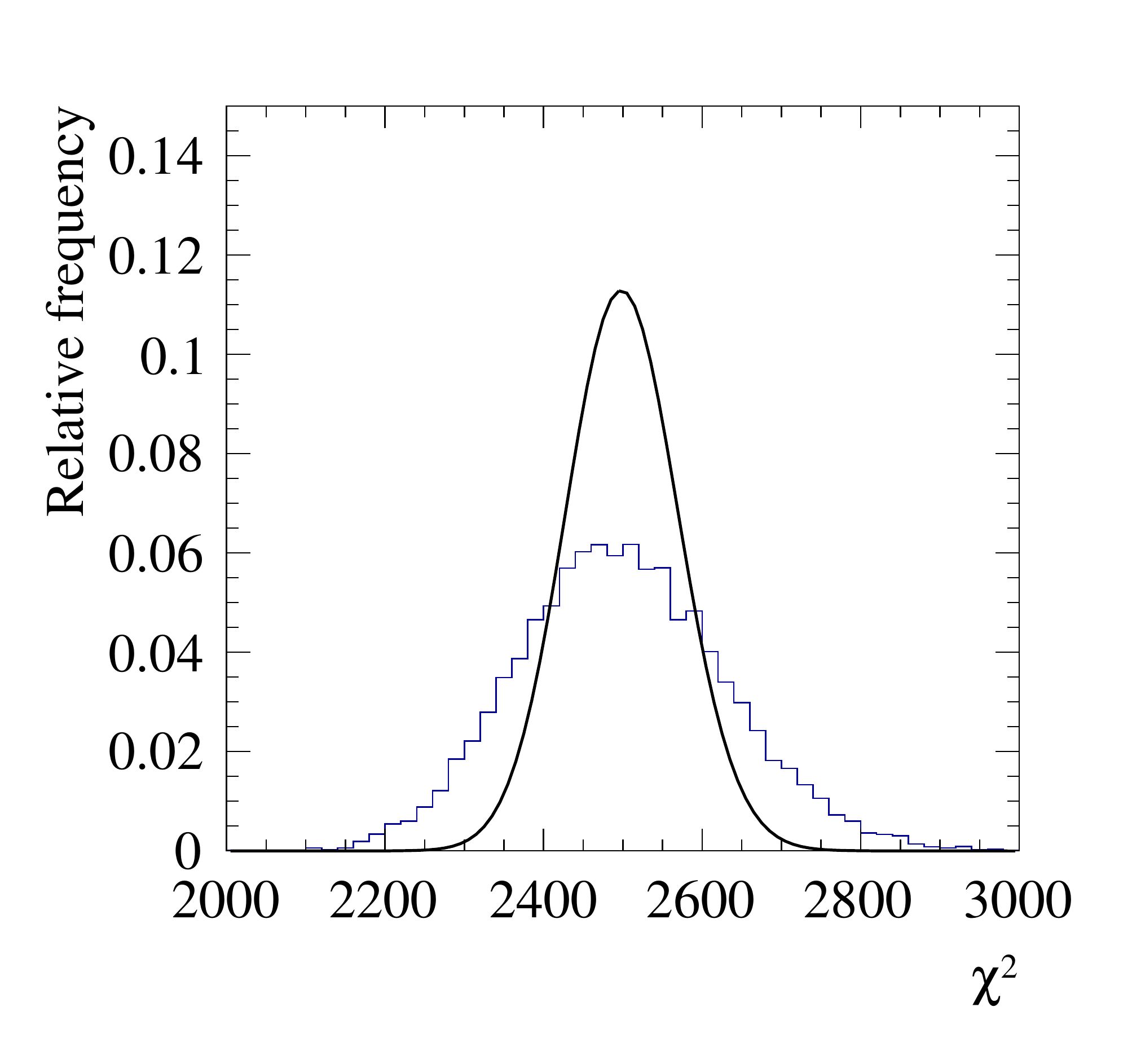}
	\caption{Observed (histogram) and expected $\chi^2$ distributions (solid line) calculated in the region $(x,y)\in [-30,30]^2$.}
	\label{fig:ex1ChiSqrReal}
\end{figure}

As a test for the convergence of the Markov Chain, the phase space is
divided in Cartesian coordinates into $50 \times 50$ bins in a region
$[-30,30]$ in $x$ and $y$. The distribution of the $\chi^{2}$ defined
in Equation~\ref{eqn:chi2} is obtained by scaling the target function
to the predefined number of samples, $N$, and by generating ensembles
of two-dimensional histograms for which each bin is a random number
drawn from a Poisson distribution around the expectation value of the
scaled target function. The distribution of the $\chi^{2}$ variable is
shown in Figure~\ref{fig:ex1ChiSqrReal}. It has a mean value of 2,500
and is wider than the expected $\chi^{2}$-distribution due to
non-Gaussian fluctuations in the low-probability region of the target
function. 

\begin{figure}[!ht]
	\centering
	\includegraphics[width=0.49\textwidth]{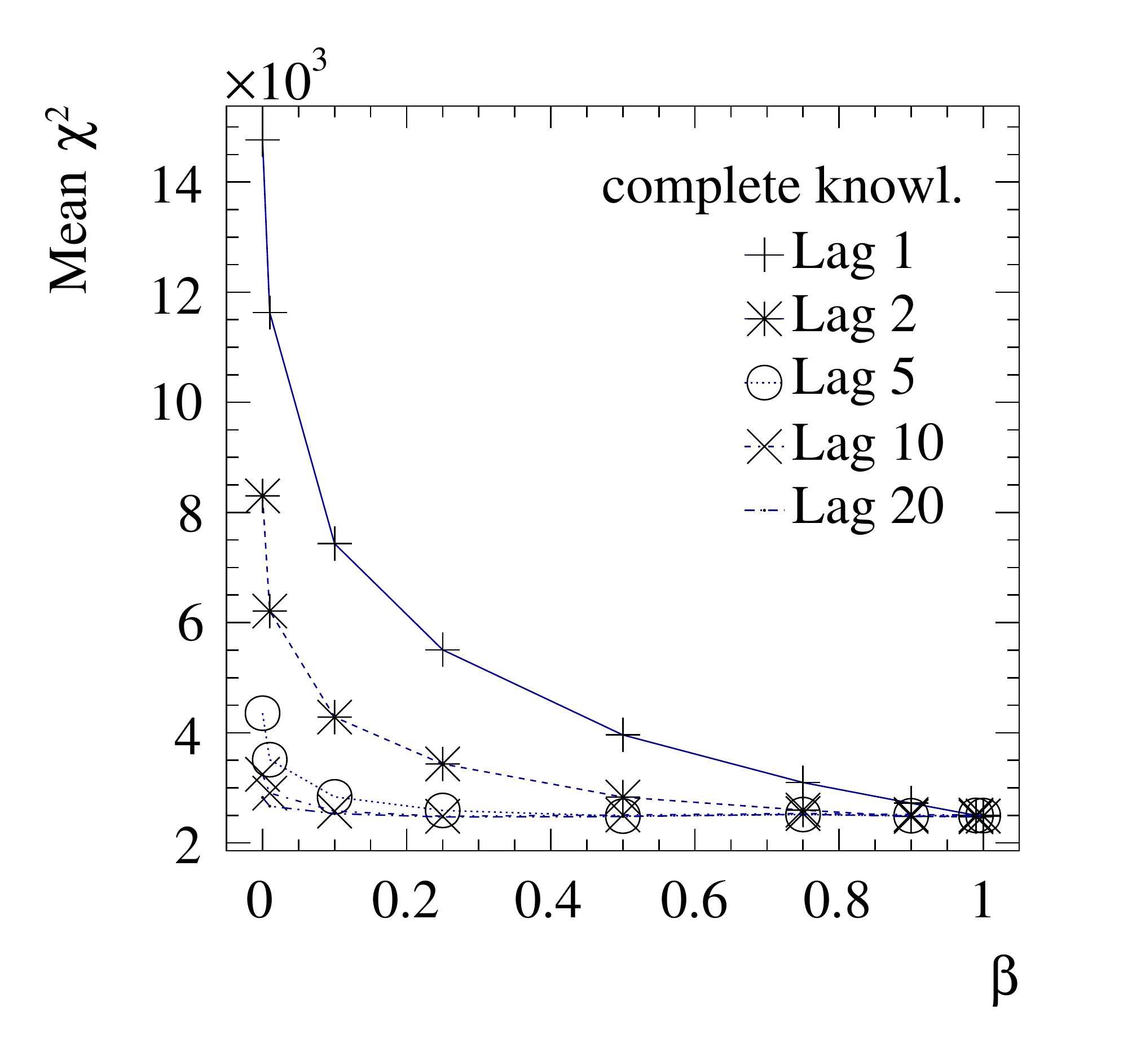}
	\includegraphics[width=0.49\textwidth]{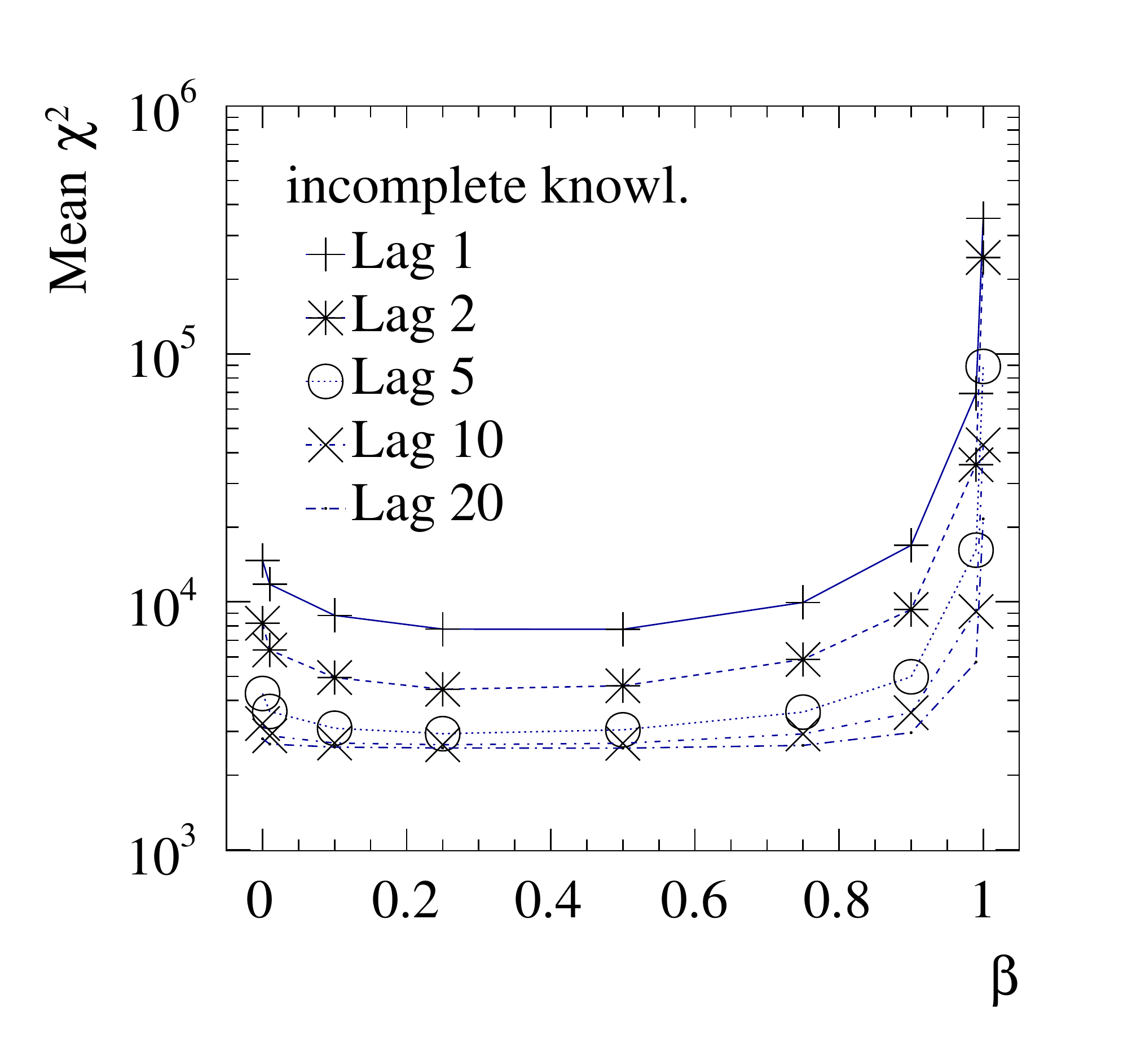}
	\caption{The mean $\chi^2$ as a function of $\beta$ for different lags for the case of complete (left) and incomplete prior knowledge (right).}
	\label{fig:ex1ChiSqr}
\end{figure}

Figure \ref{fig:ex1ChiSqr} shows the mean of the $\chi^2$ distribution
obtained from 100 runs of the \mccube\ algorithm as a function of
$\beta$ for different lags. In the case of complete prior knowledge
and a lag of one, the $\chi^{2}$ drops from roughly $15,000$ at
$\beta=0$ to about $2,500$ at $\beta=1$. The behaviour for small
values of $\beta$ is expected due to a large amount of autocorrelation
between the samples if the MH transition kernel dominates. For large
values of $\beta$ it is expected that a perfect prior knowledge leads
to a fast convergence of the Markov Chain. The trend is similar for
larger values of the lag where the mean $\chi^{2}$ converges to a
constant value of about $2,500$ rather quickly.

In the case of incomplete knowledge the mean $\chi^{2}$ value is
larger for $\beta=1$ compared to $\beta=0$, and the curve shows a
minimum for a mixed transition kernel. The large $\chi^{2}$ values
around $\beta=1$ can be explained by the fact that the proposal
function does not sample the line segment of the target function
homogeneously. The occurrence of a minimum in the curve shows that the
sampling improves compared to a pure Metropolis--Hastings algorithm if
prior knowledge is used, and that it cures the problem of incomplete
knowledge for a pure IS algorithm. The minimal $\chi^{2}$ converges to
$2,500$ with increasing lag.

\subsection{Example two: a toy generator for Drell-Yan events}
The second example represents an application of the \mccube\ algorithm
we aim for in the future, namely the generation of unweighted
events. The process under study is Drell-Yan production of lepton
pairs in proton-proton collisions at an assumed centre-of-mass energy
of 8\,TeV. The target function is defined by the differential cross
section in the partonic invariant mass squared of the lepton pair,
$s$, the rapidity of the lepton system, $Y$, the scattering angle
of the centre-of-mass system, $\theta$, and the azimuthal angle,
$\phi$. The flavour of the incoming quark is fixed to the up-quark
and the phase space is constrained to $\sqrt{s} \in [15, 200]$\,GeV, $Y \in
[-6,6]$, $\cos{\theta} \in [-1,1]$ and $\phi \in [0,2\pi]$.

The mapping function is a sum of two mappings, namely the photon and
$Z$-boson contributions, providing a reasonable approximation of the
differential cross section. The remaining differences between the mapping 
function and the target function are small and caused by the interference 
of the two processes. Both mappings are uniform in $\phi$, sample from 
an optimized histogram in $Y$, and both encode the information about 
$\theta$ using a functional form of $1+\cos(\theta)^{2}$. The mapping 
function of the photon contribution follows $1/s^{2}$ while that of 
the $Z$-boson exchange is parametrized by a Breit--Wigner distribution 
characterized by the mass and decay width of the $Z$-boson. The channel 
weights are chosen as $\alpha_{\gamma} = 0.76$ and $\alpha_{Z} = 0.24$.


\subsubsection{Characteristic numbers for the performance}

\begin{figure}
  \centering
  \includegraphics[width=0.49\textwidth]{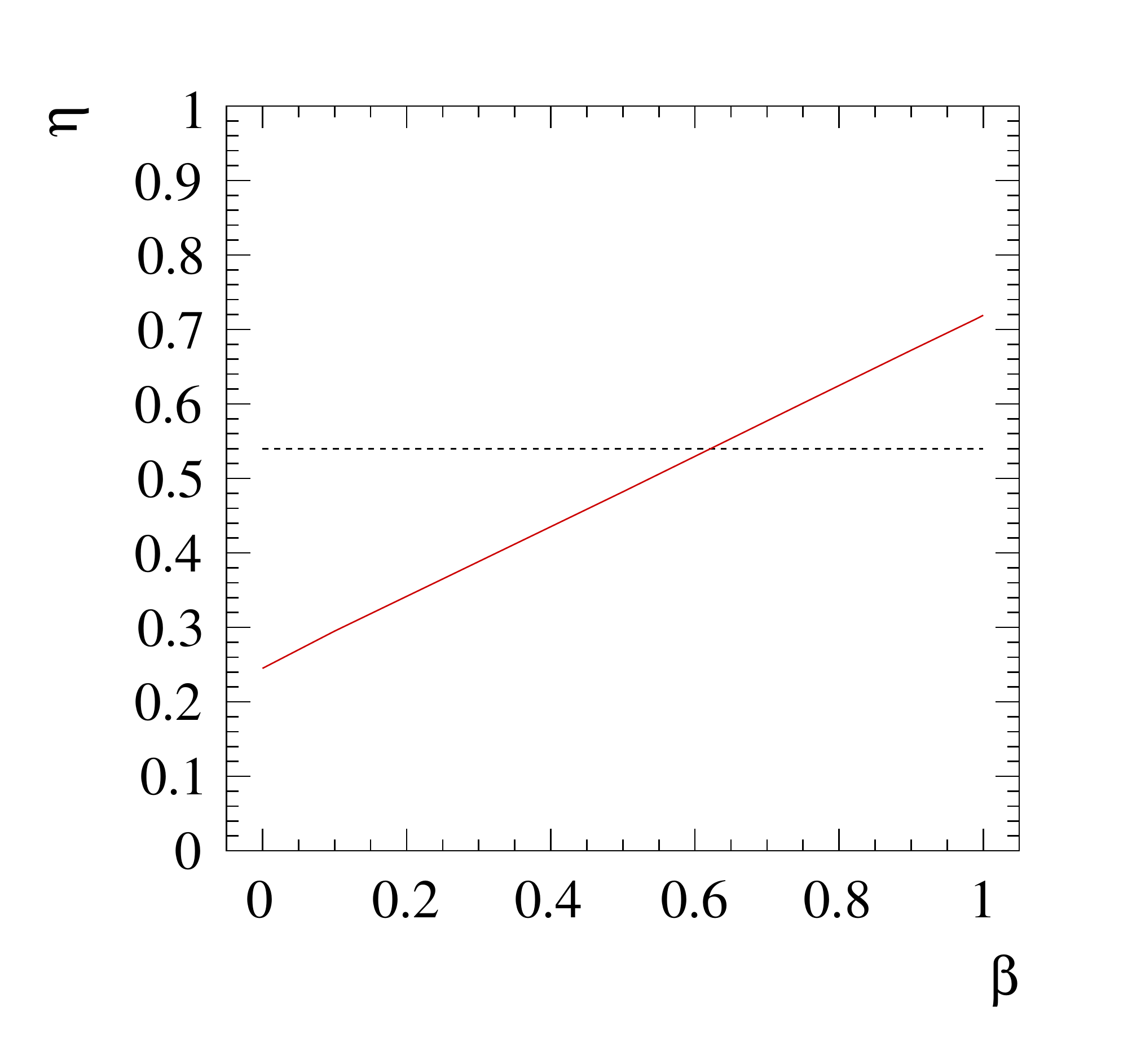}
  \includegraphics[width=0.49\textwidth]{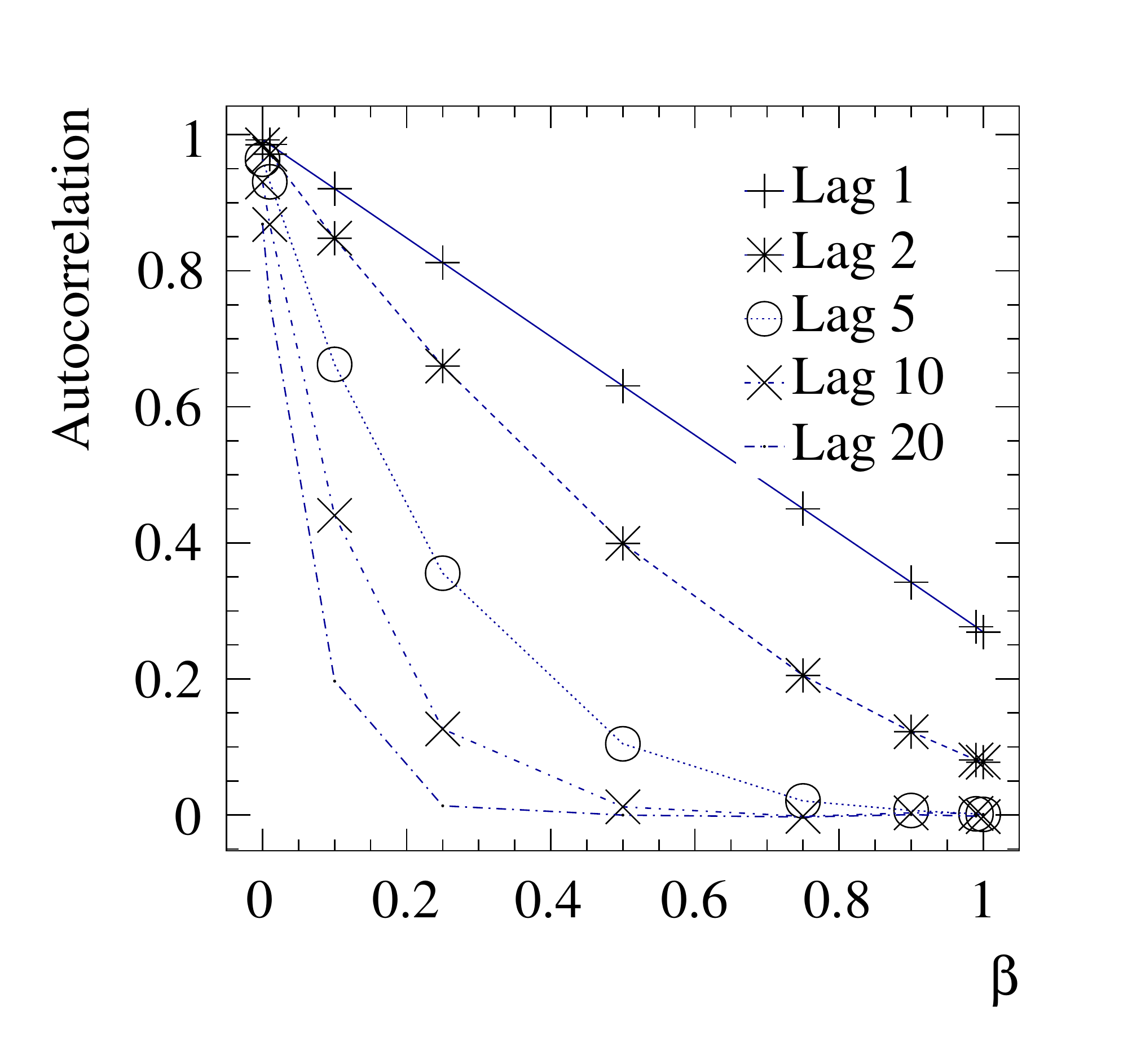}
  \caption{Left: The sampling efficiency as a function of $\beta$ for
  the \mccube\ algorithm. The sampling efficiency for the IS algorithm
  is indicated as a dashed line. Right: Autocorrelation for $\sqrt{s}$ as a
  function of $\beta$ for different lags.}
  \label{fig:ex2eff}
\end{figure}

Figure~\ref{fig:ex2eff} (left) shows the sampling efficiency for
the \mccube\ algorithm as a function of $\beta$. The efficiency
increases linearly from 25\% at $\beta=0$ to about 70\% at
$\beta=1$. In comparison, the sampling efficiency for the IS algorithm
is about 55\%.

\begin{figure}
  \centering
  \includegraphics[width=0.49\textwidth]{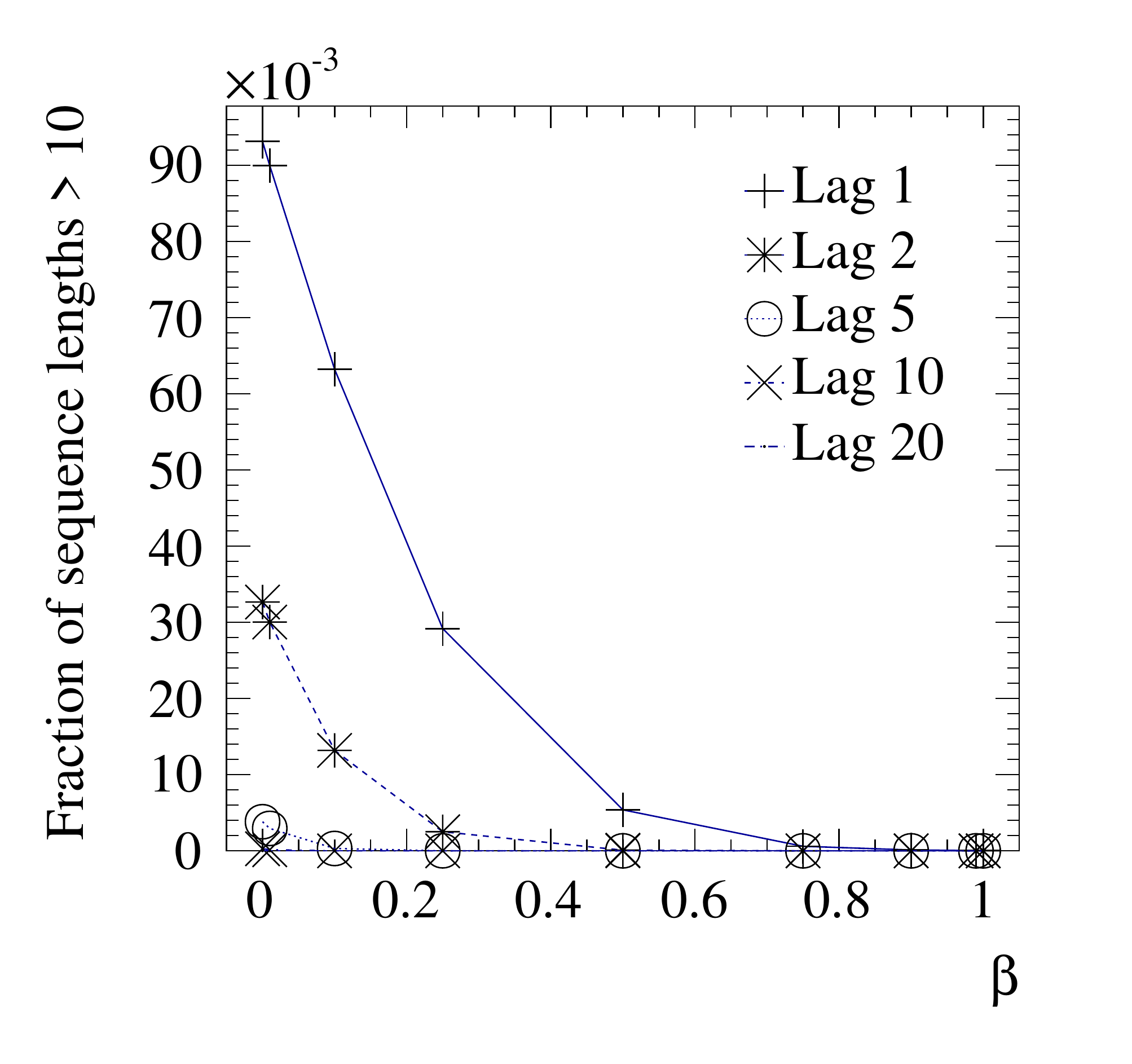}
  \includegraphics[width=0.49\textwidth]{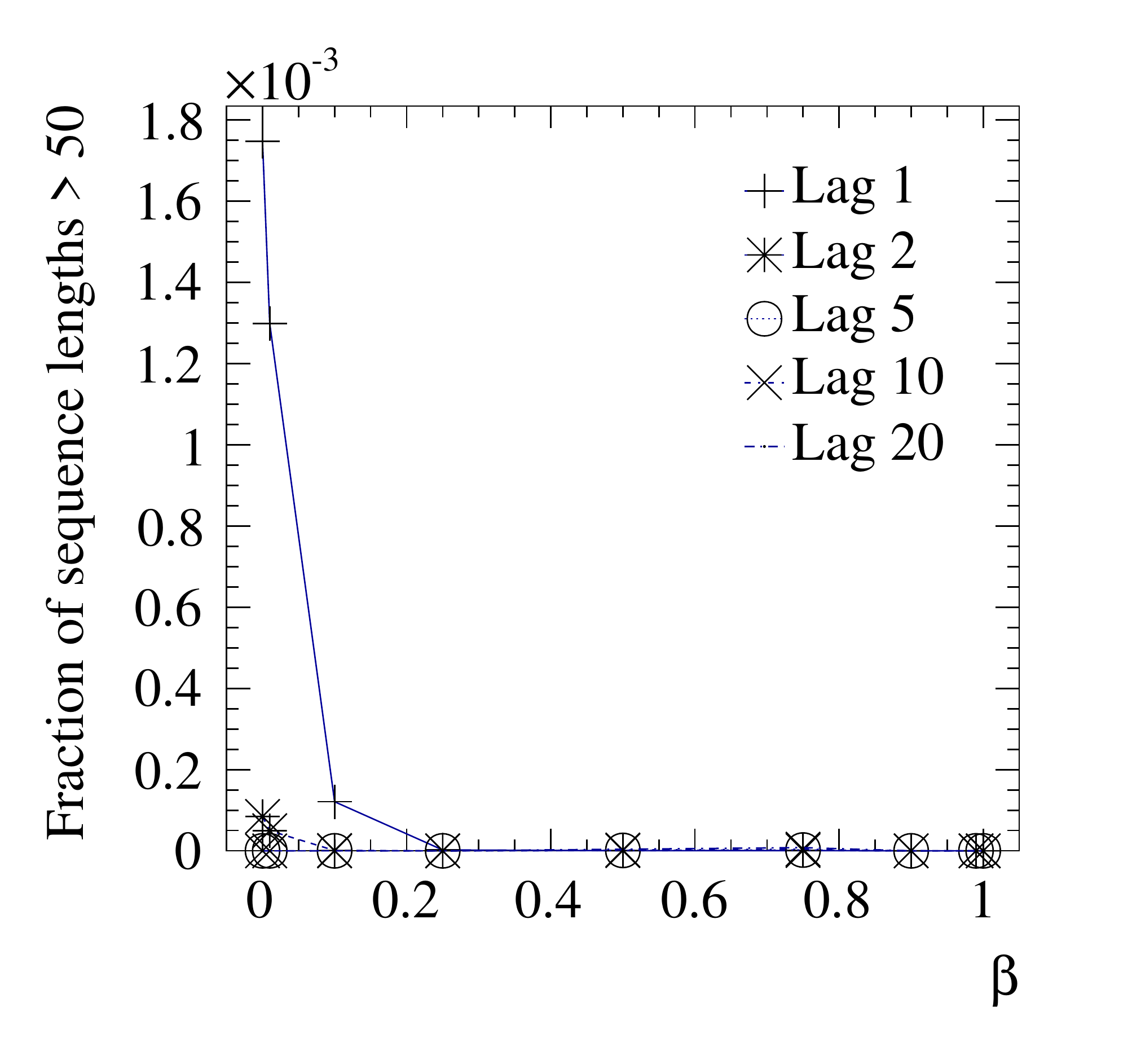}
  \caption{The fraction of sequence lengths greater than 10
  (left) and 50 (right) shown for lags between 1 and 20.}
  \label{fig:ex2seq}
\end{figure}

An opposite trend can be observed for the autocorrelation in the same
figure (right). It becomes smaller with increasing $\beta$ similar to
the behaviour in the first example. The autocorrelation ranges from
values of about 95\% at $\beta=0$ to roughly 25\% at $\beta=1$ for a
lag of one. For larger values of the lag, the autocorrelation drops
exponentially and vanishes at $\beta=1$ for lags greater than
five. This observation is consistent with the fraction of sequence
lengths greater than 10 and 50 as a function of $\beta$ which is shown
in Figure~\ref{fig:ex2seq}. Both fractions decrease with increasing
values of $\beta$ and for large lags.

\begin{figure}
\centering
\includegraphics[width=0.49\textwidth]{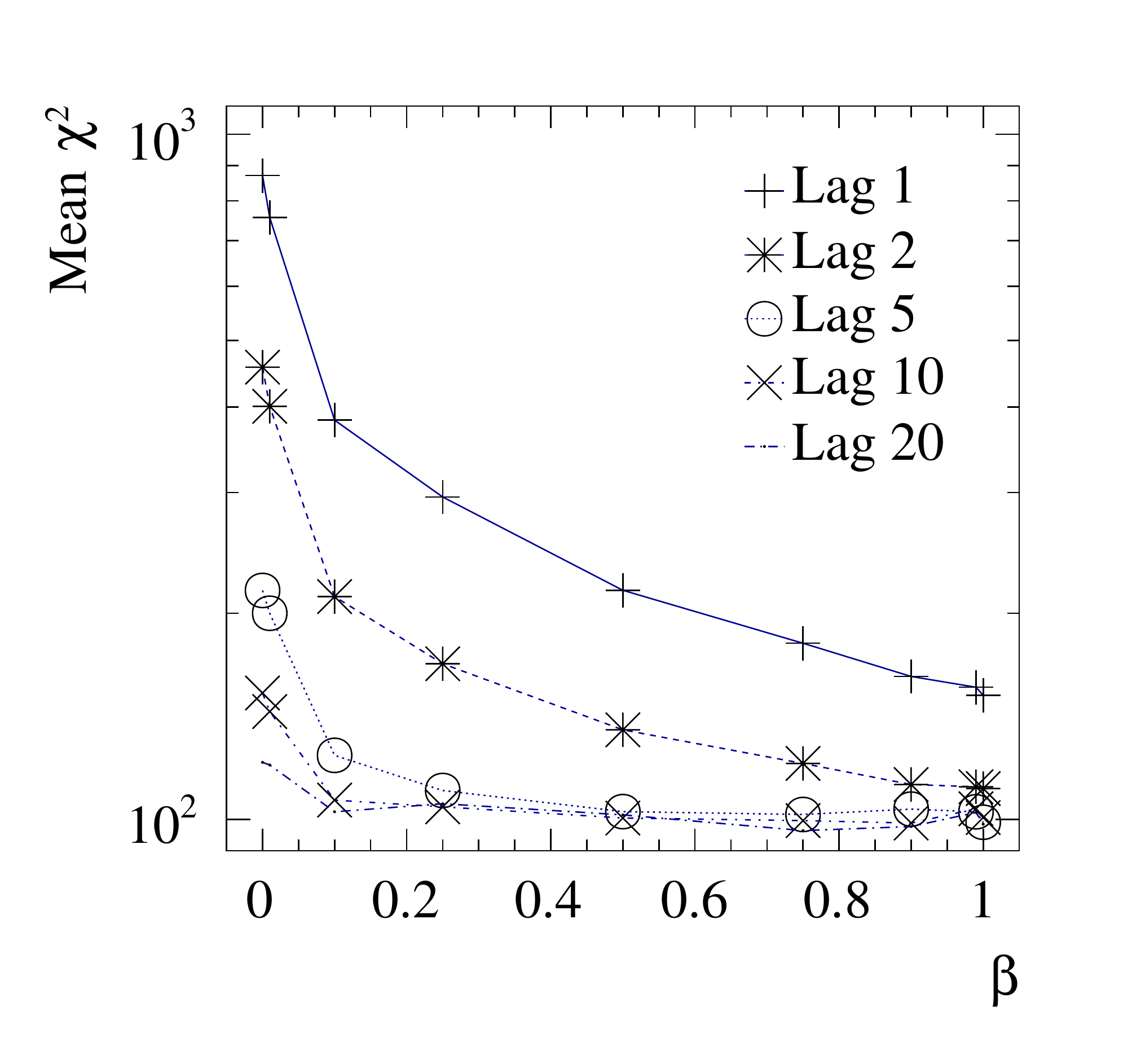}
\caption{The mean $\chi^{2}$ as a function of $\beta$ for different lags.}
\label{fig:ex2chi2}
\end{figure}

The convergence of the Markov Chains is tested with a $\chi^{2}$ in
the $\sqrt{s}$-variable. It is shown as a function of $\beta$ for
different lags in Figure~\ref{fig:ex2chi2}. The mean $\chi^{2}$ drops
exponentially for increasing values of $\beta$. It decreases faster
for an increasing choice of lag due to the reduced autocorrelation,
and converges to the expected value of $100$.

When considering lags larger than one in this simple example, the
overall efficiency of pure Multi-Channel Importance Sampling is still
higher than that of \mccube. However, in more complicated scenarios,
e.g.  for multi-particle final states with non-trivial phase-space
cuts, one is typically confronted with unweighting efficiencies of the
order of one percent or smaller when using Importance Sampling Monte
Carlo, and we foresee a huge potential to improve these cases with our
new algorithm.

\subsection{Example three: $Z$ plus multijet production within \Sherpa}
The third example extends upon the former one, aiming for a validation
of the proposed method for a realistic and high-dimensional
problem. We study the production of $Z$ bosons associated with $n$
additional jets at the LHC with a centre-of-mass energy of $\sqrt{s} =
8\,\textup{TeV}$, and $n \in \{0,1,2,3,4\}$.  We consider unweighted
event generation according to the corresponding multi-parton
tree-level matrix elements. The $Z$ boson is required to decay into a
charged lepton pair and we constrain ourselves to diagrams involving
exactly one electroweak propagator only. For this study we have
implemented the \mccube{} algorithm within the \Sherpa event generator
framework. The channels and mappings for the importance sampling
kernel are obtained from the Multi-Channel Importance Sampler
of \Sherpa, more specifically the {\tt AMEGIC}
generator~\cite{Krauss:2001iv}. Local variations of the phase-space
points steered by the MH kernel are generated using the BAT
framework \cite{Caldwell:2008fw}. For that the kinematic phase-space
configuration of the outgoing particles of the $2 \rightarrow N$
scattering process is mapped on $3N-4$ random variables with a
parametrisation similar to the one presented in
Ref.~\cite{Platzer:2013esa}. Technical details will be provided in a future
publication.

To avoid physical singularities in the processes under consideration we need 
to regulate contributions from massless photon exchange and soft- and collinear 
QCD emissions. This is achieved by applying the following set of standard cuts:
\begin{itemize}
\item transverse momentum of the charged leptons $p_{T,\ell} \geq 20\,\textup{GeV}$;
\item invariant mass of the lepton pair $66\,\textup{GeV} \leq m_{\ell \ell} \leq 116\,\textup{GeV}$;
\item exactly $n$ anti-$k_T$ jets with transverse momenta 
$p_{T,j} \geq 30 \, \textup{GeV}$ and distance parameter $R=0.4$.
\end{itemize}

During an initial prerun the channel weights of the importance sampling
kernel are optimized such that the variance of the integral estimate, 
here the total cross section, is reduced. Then, for a kernel mixing parameter 
of $\beta=0.1$, the proposal width for each of the $3N-4$ parameters is adapted 
separately to yield a sampling efficiency between $0.35$ and $0.55$, 
respectively.

For each jet multiplicity we generated samples of unweighted 
events using both \Sherpa's standard IS algorithm and the new 
\mccube\ sampler for a kernel mixing parameter of $\beta=0.8$. 
Table \ref{tab:ex3SamplingEff} compares the respective sampling 
efficiencies as a function of jet multiplicity $n$, based on a lag 
of 1 for the latter. For the original importance sampling approach 
used in \Sherpa the sampling efficiency decreases significantly with 
an increasing number of final-state jets. 

\begin{table}
	\centering
	\begin{tabular}{| l | c | c | c | c | c |}
		\hline
		$n$\,jets & 0	& 1 &	2 &	3 &	4 \\
		\hline
		IS & $0.20$ & $6.8 \cdot 10^{-3}$ & $3.1 \cdot 10^{-3}$ & $4.0 \cdot 10^{-4}$ & $2.9 \cdot 10^{-6}$ \\
		\hline
		\mccube{} $\beta=0.8$ & $0.50$ & $0.28$ & $0.25$ & $0.19$ & $0.11$ \\
		\hline
	\end{tabular}
	\caption{Sampling efficiency for $Z+n$\,jets event generation with the \mccube{} 
                algorithm as implemented in \Sherpa in comparison to pure 
		Multi-Channel Importance Sampling (IS). A lag of 1 is used and the 
                kernel mixing parameter is fixed to $\beta=0.8$.}
	\label{tab:ex3SamplingEff}
\end{table}

With increasing final-state multiplicity the number of sub-processes
as well as the number of Feynman diagrams, i.e. phase-space topologies
and corresponding integration channels, per sub-process increases rapidly. 
This results in the significant drop of sampling efficiency for the original 
pure Multi-Channel Importance Sampler.  Furthermore, with increasing jet 
multiplicity the computational costs for the evaluation of the matrix 
element per phase-space point rise rapidly. 

In Table \ref{tab:ex3SamplingEff2} the scaling behaviour of the sampling 
efficiency as a function of the kernel mixing parameter $\beta$ is presented 
for the process $Z+3$\,jets. The expected linear decrease of $\eta$ with 
increasing values of $\beta$ is confirmed.

\begin{table}
	\centering
	\begin{tabular}{| l | c | c | c | c |}
		\hline
		$\beta$ & 0.6 & 0.7 & 0.8 & 0.9 \\
		\hline
		$\eta$ & 0.26 & 0.23 & 0.19 & 0.16 \\
		\hline
	\end{tabular}
	\caption{Sampling efficiency $\eta$ for $Z+3$\,jets production 
                for the \mccube{} implementation in \Sherpa for 
                different values of the mixing parameter $\beta$ using a
                fixed lag of 1.}
	\label{tab:ex3SamplingEff2}
\end{table}

Seemingly the \mccube{} approach outperforms pure Multi-Channel
Importance Sampling by several orders of magnitude, in particular for
high jet multiplicities. However, these improved sampling efficiencies
have to be corrected by a lag in order to account for autocorrelation
effects in the generated samples introduced by the design of
the \mccube{} algorithm.

\begin{figure}
	\begin{center}
	\includegraphics[width=0.5\textwidth]{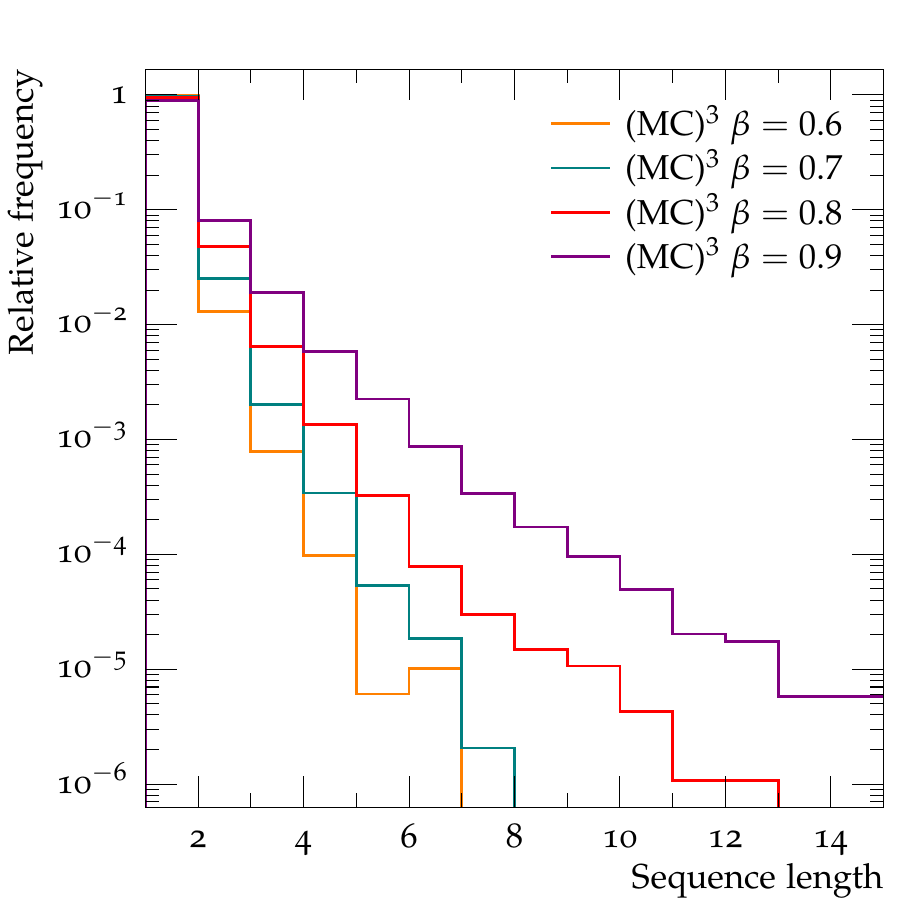}        
	\caption{Sequence lengths distribution in $Z+3$\,jets                
		production using the \mccube\ algorithm with different values of 
		the kernel mixing parameter $\beta$. 
		}
	       \label{fig:ex3SeqLen}
	\end{center}
\end{figure}

For the analysis of the statistical properties of the samples
generated with \mccube{} we consider the case of $Z+3$\,jets
production. To reduce the autocorrelation in the \mccube{} samples an
initial lag of 20 is applied during event generation. Introducing a
lag in \mccube{}, the remaining sequence lengths in the generated
samples decrease. Figure~\ref{fig:ex3SeqLen} illustrates that the
fractions of larger sequence lengths in the samples decrease indeed
exponentially. With the initial lag of 20 during production, no
sequence length above 15 is observed for any of the considered choices
for $\beta$ and considering a sample of size 1M events. This simple
measure of the autocorrelation increases with an increasing mixing
parameter $\beta$. Note that the efficiencies quoted earlier have to
be corrected for the lag. For the chosen working point, they still
show a significant improvement over those obtained from a pure IS
algorithm for large jet multiplicities.

\begin{figure}
	\includegraphics[width=0.5\textwidth]{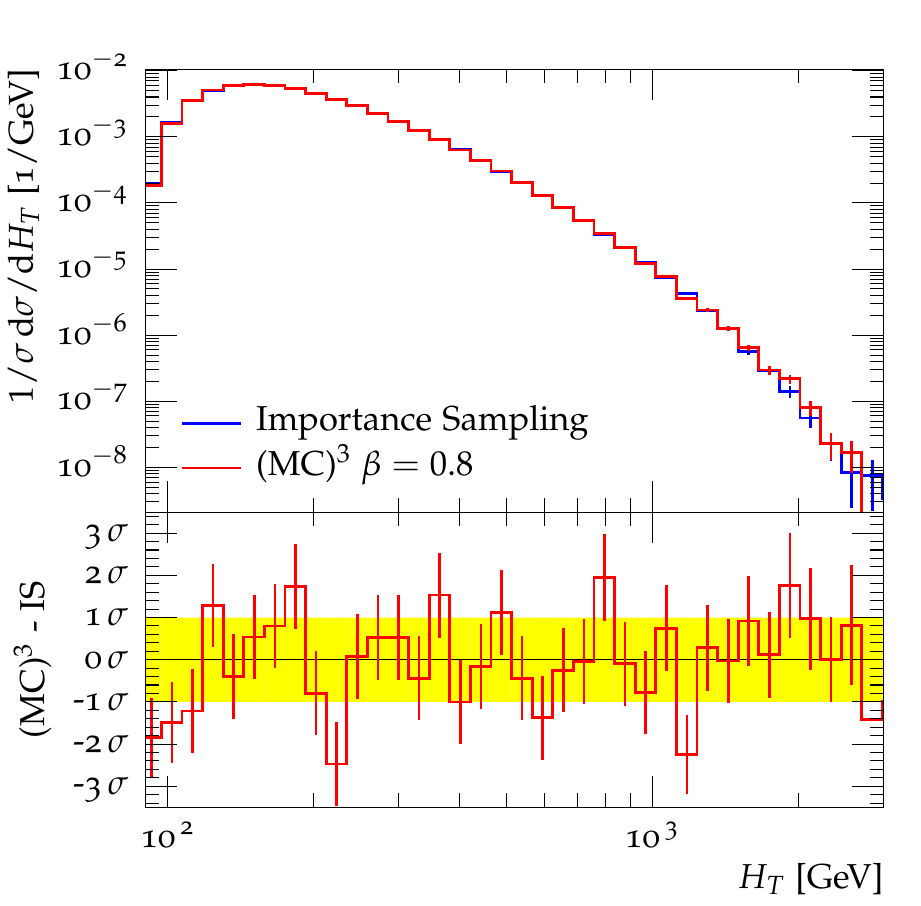}
	\includegraphics[width=0.5\textwidth]{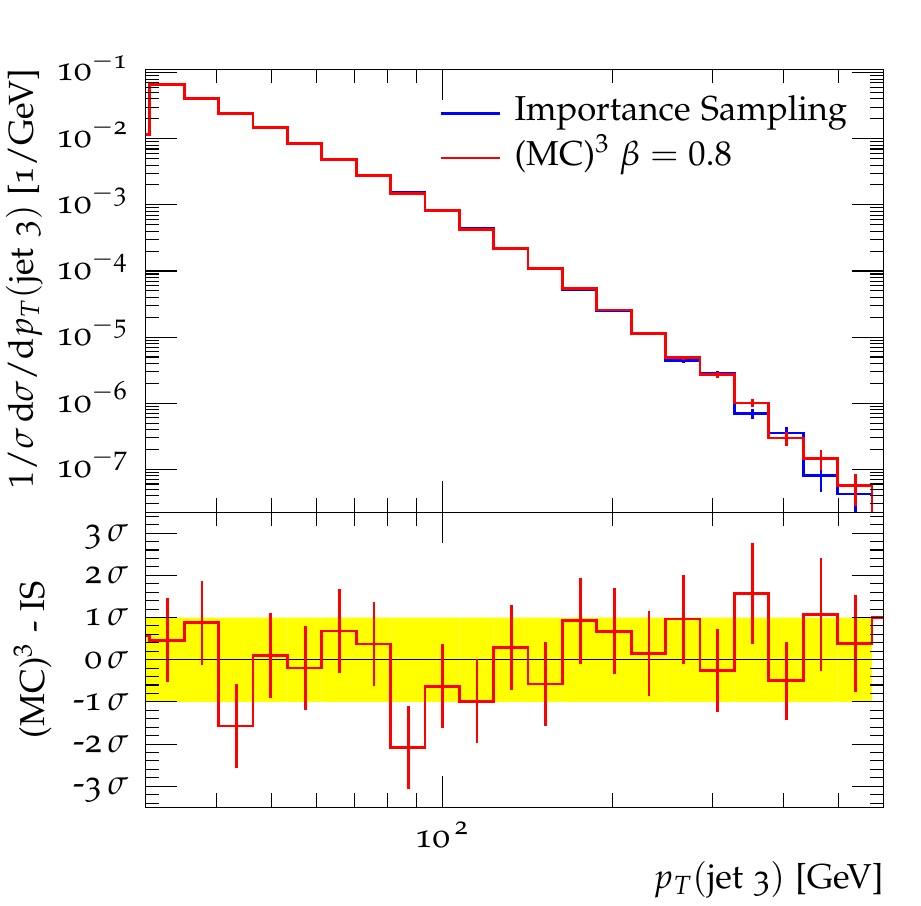}
	\caption{Normalized differential distributions for the third-jet 
                transverse momentum $p_T$ (right) and the scalar sum of 
                all jet transverse momenta $H_T$ (left) in $Z+3$\,jets 
                production. Shown are the predictions from pure 
                Multi-Channel Importance Sampling (blue) and the 
                \mccube\ algorithm with a lag of 20 and $\beta=0.8$. 
                The lower panels indicate the statistical compatibility
                of the samples, measured in terms of standard deviations
                of the IS result. 
		}
	\label{fig:ex3StatCom}
\end{figure}

We close our discussion of this example by testing the consistency of
observable distributions predicted by the two algorithms for the 
$Z+3$\,jets process. In Figure~\ref{fig:ex3StatCom} we analyse the
statistical compatibility of the sample generated with \mccube{} using
a lag of 20 and a kernel mixing parameter $\beta=0.8$ with the
reference sample generated using the original Multi-Channel Importance
Sampling approach. We present results for the transverse momentum of
the third jet, $p_T({\rm jet}\,3)$, and the scalar sum of all jet
transverse momenta, $H_T$. As a measure for the statistical compatibility
of the two samples we indicate in the lower panels the bin-wise
difference measured in terms of standard deviations of the IS result.
Clearly both approaches yield fully compatible results, not only for
the observables presented here, allowing us to conclude that the
\mccube\ algorithm yields a fully consistent event generation routine
that can clearly supersede standard importance sampling methods in 
particular for high-multiplicity final states. 

\section{Conclusions}
\label{sec:conclusions}
We have presented a new algorithm for phase-space sampling called 
\mccube. It improves pure Markov Chain Monte Carlo techniques by 
incorporating prior knowledge about the target function from a
corresponding Multi-Channel Importance Sampling algorithm. \mccube\
makes use of a linearly mixed transition kernel given by a locally 
acting Metropolis--Hastings component and an importance sampling 
kernel that allows for global jumps in phase space.  

We have assessed the systematics of the new algorithm with three
illustrative examples, thereby focusing on the sampling probability,
autocorrelation effects and the convergence of the resulting Markov
Chain. We have shown that incomplete prior knowledge can cause a
severe drop in sampling efficiency when using Multi-Channel Importance
Sampling, and thus an increased number of calls to the target
function.  Even for problems with a low number of dimensions, this can
be particularly severe if resonant structures in the mapping function
$g$ are missing. In contrast, the \mccube\ algorithm can increase the
sampling efficiency because of the self-adapting properties of the
produced Markov Chains.  However, the resulting samples show an
autocorrelation, and its strength depends on the amount of prior
knowledge. The latter is controlled by the parameter $\beta$ and the
lag. The impact of the autocorrelation can also be seen in the
distribution of the sequence lengths and discrepancy variables which
show differences between the true and the sampled distribution. A
large autocorrelation also indicates a poor convergence of the Markov
Chain to its limiting distribution. The autocorrelation can be suppressed 
to a reasonable level by choosing a small to moderate lag.

We have shown that the \mccube\ algorithm works very well for a low
number of dimensions and that it performs better than the traditional
Multi-Channel Importance Sampling for the case of incomplete prior
knowledge. The third example also shows that the new algorithm
outperforms the Importance Sampling algorithm for an increasing number
of final-state particles when applied to the concrete task of
producing unweighted events with Monte Carlo event generators.


\section*{Acknowledgements}
\label{app:acknowledgements}
We wish to thank Allen Caldwell, Frederik Beaujean, Daniel
Greenwald and Andre van Hameren for the useful discussions and their 
feedback.


\bibliographystyle{elsarticle-num}
\biboptions{sort&compress}
\bibliography{refs}

\end{document}